\documentclass[12pt,preprint2]{aastex}

\usepackage{lscape}
\shorttitle{Orbital motion of the Quintuplet cluster}
\shortauthors{A. Stolte et al.}

\begin{document}

\title{The orbital motion of the Quintuplet cluster - a common origin for 
the Arches and Quintuplet clusters?$^\ast$}
\author{A. Stolte\altaffilmark{1}, B. Hu{\ss}mann\altaffilmark{1}, 
M. R. Morris\altaffilmark{2}, A. M. Ghez\altaffilmark{2}, W. Brandner\altaffilmark{3}, 
J. R. Lu\altaffilmark{4}, W. I. Clarkson\altaffilmark{5}, M. Habibi\altaffilmark{1}, 
K. Matthews\altaffilmark{6}}
\altaffiltext{1}{Argelander Institut f\"ur Astronomie, Auf dem H\"ugel 71, 53121 Bonn, Germany, astolte@astro.uni-bonn.de}
\altaffiltext{2}{Division of Astronomy and Astrophysics, UCLA, Los Angeles, CA 90095-1547, ghez@astro.ucla.edu, morris@astro.ucla.edu}
\altaffiltext{3}{Max-Planck-Institut f\"ur Astronomie, K\"onigstuhl 17, 69117 Heidelberg, Germany, brandner@mpia.de}
\altaffiltext{4}{Institute for Astronomy, University of Hawai'i, 2680 Woodlawn Drive, Honolulu, HI 96822, jlu@ifa.hawaii.edu}
\altaffiltext{5}{Department of Natural Sciences, University of Michigan-Dearborn, 125 Science Building,
4901 Evergreen Road, Dearborn, MI 48128, wiclarks@umich.edu}
\altaffiltext{6}{Caltech Optical Observatories, California Institute of Technology, MS 320-47, Pasadena, CA 91225, kym@caltech.edu}
\altaffiltext{*}{This paper is based on observations 
obtained with the VLT at Paranal Observatory, Chile, under programme
71.C-0344 (PI: Eisenhauer) and 081.D-0572(B) (PI: Brandner), 
and obtained with the Keck telescopes at Mauna Kea, Hawai'i (PI: Morris).} 

\begin{abstract}
We investigate the orbital motion of the Quintuplet cluster near the Galactic center
with the aim of constraining formation scenarios of young, massive star clusters in nuclear
environments.
Three epochs of adaptive optics high-angular resolution imaging with the 
Keck/NIRC2 and VLT/NAOS-CONICA systems were 
obtained over a time baseline of 5.8 years, delivering an astrometric accuracy of 0.5-1 mas/yr.
Proper motions were derived in the cluster reference frame and were used to distinguish cluster 
members from the majority of the dense field star population toward the inner bulge.
Fitting the cluster and field proper motion distributions with 2D gaussian models,
we derive the orbital motion of the cluster for the first time. The Quintuplet
is moving with a 2D velocity of $132 \pm 15$ km/s with respect to the field 
along the Galactic plane, which yields a 3D
orbital velocity of $167 \pm 15$ km/s when combined with the previously known radial velocity.
From a sample of 119 stars measured in three epochs, we derive an upper limit to the 
velocity dispersion in the core of the Quintuplet cluster of $\sigma_{1D} < 10$ km/s.
Knowledge of the three velocity components of the Quintuplet allows us to
model the cluster orbit in the potential of the inner Galaxy. 
Under the assumption that the Quintuplet is located in the central 200 pc at the present time, 
these simulations exclude the possibility that the cluster is moving on a circular orbit.
Comparing the 
Quintuplet's orbit with our earlier measurements of the Arches orbit, we discuss the possibility
that both clusters originated in the same area of the central molecular zone. 
According to the model of Binney et al.~(1991), two families of stable cloud orbits are located 
along the major and minor axes of the Galactic bar, named x1 and x2 orbits, respectively.
The formation locus of these clusters is consistent with the outermost x2 orbit 
and might hint at cloud collisions at the transition 
region between the x1 and x2 orbital families located 
at the tip of the minor axis of the Galactic bar. The formation of young, massive star clusters
in circumnuclear rings is discussed in the framework of the channeling in of dense gas by the
bar potential. We conclude that the existence of a large-scale bar plays a major role in supporting
ongoing star and cluster formation, not only in nearby spiral galaxies with circumnuclear rings, 
but also in the Milky Way's central molecular zone.
\end{abstract}


\section{Introduction}

The Galactic center is host to 3 young, massive star clusters,
the Arches and Quintuplet clusters at projected distances of 26 to 32 pc
from the supermassive black hole (Nagata et al.~1990, 1995, Okuda et al.~1990, 
Cotera et al.~1996, Figer et al.~1999ab, Stolte et al.~2008), 
and the young nuclear cluster inside the central few parsecs 
(e.g., Genzel et al.~2003, Ghez et al.~2005, Paumard et al.~2006, 
Sch\"odel et al.~2007, Do et al.~2009, 2013, Bartko et al.~2010, Lu et al.~2013). 
Each of these clusters is known to 
host at least $10^4\,M_\odot$ in stars, with the Arches and Quintuplet
likely containing stellar masses in excess of $2\times 10^4\,M_\odot$
(Figer et al.~1999a, Espinoza et al.~2009, Clarkson et al.~2012, Habibi et al.~2013),
similar to the stellar content in the Young Nuclear Cluster (Bartko et al.~2010, Lu et al.~2013).
At ages of only a few Myr, these clusters harbour a rich population 
of Wolf-Rayet stars and supergiants, which are the 
youngest and most extreme post-main sequence evolutionary phases 
of O-type stars (Crowther 2007, Martins et al.~2008).  
The Quintuplet cluster in particular is associated with at least two sources
in the short-lived Luminous Blue Variable (LBV) phase and contains several carbon-enriched
Wolf-Rayet (WC) stars (Figer et al.~1999b, Liermann et al.~2009, 2010, Mauerhan et al.~2010a). 
The short dynamical lifetime of young clusters in the GC
of only a few 10 Myr (Kim et al.~2000, Portegies Zwart et al.~2002)
suggests that the Arches and Quintuplet clusters contribute
to the apparently isolated field population of evolved, high-mass stars 
(Mauerhan et al.~2010b). 

At the same time, the origin of these massive clusters is unknown. 
The dynamical properties of a young star cluster are indicators of
the cluster's origin and stability during its evolutionary timescale. 
In contrast to spiral arm clusters, clusters emerging near the center of the Galaxy 
are moving rapidly from their birth sites and do not appear affiliated with their
natal clouds. The orbital motion of these clusters, combined with the cluster age, 
provides the only clue to their birth sites by tracing the cluster orbit backwards 
in time. In addition, the 
orbital velocity is an important determinant for the chances of cluster survival. 
Clusters on orbits very close to the Galactic center experience stronger tidal losses, 
but a larger orbital velocity decreases the effect of dynamical friction in the 
central potential.

With the goal of tracing back the dynamical evolution of the Arches and Quintuplet 
clusters to their potential formation locus, we have set out to measure 
the orbital motion and the internal velocity dispersion of both clusters
from proper motions, covering a time baseline of six years. 
With the first derivation of the 3D orbital velocity
of the Arches cluster and its orbital path for a family of nested orbits at 
various line-of-sight distances (Stolte et al.~2008), we constrained
the most likely formation locus of this cluster to the inner 200 pc and hence
showed for the first time that these massive clusters might form within 
the boundaries of the central molecular zone (CMZ). One predestined location
where cloud collisions might trigger cluster formation are the regions 
where instreaming clouds on x1 orbits from the outer Galactic bar 
collide with CMZ clouds on orbits around the bar's minor axis (x2 orbits,
Binney et al.~1991). In our previous paper, we suggested a formation locus 
for the Arches cluster near the transition region of the x1 and 
x2 orbital families (Binney et al.~1991).
Recent results by the Herschel satellite have fostered the notion of a star-forming
ring in the inner 240 pc (diameter) of the CMZ (Molinari et al.~2011). 
Additional support
to cluster formation in the Galactic center stems from the discovery of dense, 
compact clouds with sufficient masses to form $10^4\,M_\odot$ clusters 
(Longmore et al.~2012, 2013). The clouds are located in projection along
the star-forming ring proposed by Molinari et al.~(2011). Although both the 
Arches and Quintuplet clusters were also suggested to be located on this 
structure (Longmore et al.~2013), their orbital velocities are higher by 
about a factor of two as compared to the terminal cloud velocities 
of $\sim 100$ km/s (Molinari et al.~2011).

In this contribution, we present the orbital velocity and obtain an upper limit
to the internal velocity dispersion of the Quintuplet cluster. 
Combining the proper motion of the Quintuplet with its radial velocity,
we model the cluster's orbital motion as a function of the line-of-sight distance, 
which provides clues on the formation locus of the cluster. 
Ultimately, the comparison with N-body 
simulations will yield the expected tidal losses that have 
occured during the cluster's lifetime.
Comparable results for the - presumably younger - Arches cluster are 
presented in Stolte et al.~(2008) and Clarkson et al.~(2012).

%%% labels changed in retrospect & send to ApJ - AS 16 June 2014 :)
% Sec. 2 \label{obssec}
% Sec. 3 \label{pmsec}
% Sec. 3.3 \label{orbitsec}
% Sec. 4 \label{discsec}
% Sec. 5 \label{sumsec}

The paper is organized as follows: The observations are described in 
Sec.~\ref{obssec}, followed by the data reduction, astrometry and 
photometry procedures. The orbital motion and velocity dispersion 
are fitted using the proper motion plane in Sec.~\ref{pmsec}, while 
simulations of the family of cluster orbits are shown in Sec.~\ref{orbitsec}. 
We also present an updated version of the Arches cluster orbit for comparison, 
adjusted for the slightly lower orbital velocity found in our new multi-epoch 
investigation (Clarkson et al.~2012). In Sec.~\ref{discsec}, we discuss the 
implications for cluster formation in the nucleus of the Milky Way and other galaxies.
Sec.~\ref{sumsec} summarizes the main findings of the proper motion analysis.

\section{Observations}
\label{obssec}

For the proper motion analysis of the central region of the Quintuplet, 
data from the ESO Very Large Telescope (VLT) 
taken in 2003 were combined with Keck observations obtained in 2008 and 2009.
A second epoch of NACO observations obtained in July 2008 was used to constrain
the 2D cluster motion from a sample of stars at larger radii from the cluster
center.
All positions are approximately centered on the central Quintuplet star Q12
(Glass et al.~1990) at RA 17:46:15.12, DEC -28:49:35.06.
A technical summary of the observations is provided in Table \ref{obstab}.

\subsection{Keck/NIRC2}
\label{keckobssec}

Keck observations were carried out using the NIRC2 near-infrared camera
(PI: K.~Matthews) behind laser guide star adaptive optics 
on 2008 May 14 and 2009 May 4. The narrow camera was used with a field 
of view of $10{''}$, delivering a pixel scale of 9.942 milliarcseconds (mas) 
per pixel. A detector integration time of 1s with 
30 co-added frames led to a total integration time of 30s per image using 
the $K'$ filter ($\lambda_c = 2.124\mu$m, $\Delta\lambda = 0.351\mu$m),
similar to the specifications of the NACO $K_s$ filter (Sec.~\ref{nacoobssec}).
A small dither pattern of $\pm 0.35{''}$ in 2008 and of $\pm 0.60{''}$ in 2009
was used to facilitate bad-pixel removal. The small dither movements 
minimize optical distortion effects, which enhances the astrometric
accuracy of the NIRC2 data set.
In the 2008 campaign, 58 images were obtained with Strehl ratios 
ranging from 8\% to 34\%, with 
a spatial resolution between 57 and 108 mas (Full Width at Half Maximum, FWHM).
As spatial resolution is the limiting factor for the astrometric accuracy,
only the 33 frames with $FWHM < 70$ mas and Strehl ratios $SR > 20$\% were 
selected for the astrometric analysis. 
In the 2009 campaign, the same exposure time setup was used to obtain 
117 images with Strehl ratios between 12 and 47\% and 
spatial resolutions of $50 < FWHM < 116$ mas. In total 60 frames 
with $SR > 28$\% and $FWHM < 62$ mas were selected for image combination. 

\begin{figure*}
\hspace*{7mm}
\includegraphics[width=16cm]{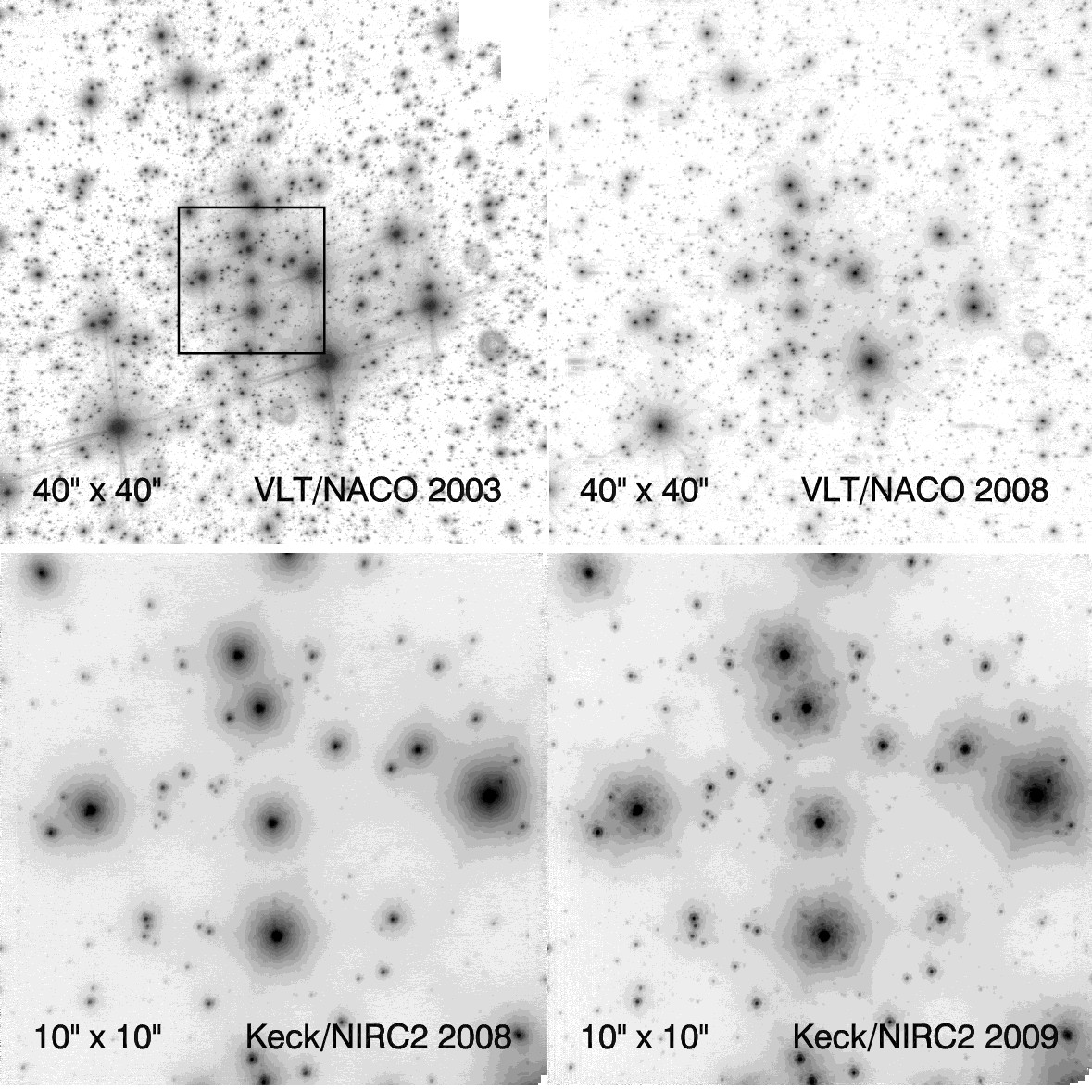}
\caption{\label{quinarch} Three proper motion epochs of the Quintuplet cluster 
with images from 2003 to 2009. 
In the {\sl top panels}, the complete $40{''} \times 40{''}$ 
dithered field of view obtained with VLT/NACO in 2003 and 2008 is displayed.
The box in the center of the 2003 image shows the region where the NACO 
astrometry was combined with the Keck/NIRC2 2008 and 2009 epochs for the 
central data set with highest astrometric accuracy.
In the {\sl bottom panels}, the Keck/NIRC2 2008 and 2009 epochs of the
central $10{''}$ or 0.4 pc of the cluster are shown.
North is up and East is to the left.}
\end{figure*}

The data reduction was carried out with our custom-made NIRC2 data pipeline
(see Lu et al.~2009 for a detailed description).
The raw data were reduced using dark images with the same exposure time, 
readout mode, and number of co-adds, and were then divided by the master 
flat-field created from on-off lamp flats taken during the same night. 
During the combination of the master dark and flat-field images, a 3 sigma 
clipping routine was applied to detect hot and dark pixels and create a fixed
NIRC2 bad pixel mask. 
In addition to the fixed NIRC2 bad pixel mask, individual mask images were 
extracted for each image using a standard cosmic ray detection routine
({\sl crrej} in IRAF).
At the end of the cluster observing sequence, 9 sky images were obtained with 
the same observational setup and at an airmass comparable to that of the science 
observations. The sky images were scaled to a common mean sky level prior to image 
combination, and the resulting master sky was scaled with scaling factors of
0.99-1.03 to the background flux of each science frame before sky subtraction. 
The reduced images were combined using the {\sl drizzle} task (Fruchter \& Hook 2002)
in IRAF\footnote{%
IRAF is distributed by the National Optical Astronomy Observatory, which is 
operated by the Association of Universities for Research in Astronomy (AURA) 
under cooperative agreement with the National Science Foundation.}. 
In addition to a deep image encompassing all 33 (60) selected frames in 2008 (2009)
shown in Fig.~\ref{quinarch}, 
3 auxiliary images were drizzled from three subsets of 11 (20) frames each, 
where the full range of spatial resolutions is included in each stack
to achieve a comparable data quality among all three subsets. Photometry is 
derived on these auxiliary images in the same way as on the deep combined image
to obtain photometric and astrometric uncertainties from repeated measurements.

\subsection{VLT/NAOS-CONICA}
\label{nacoobssec}

The Quintuplet cluster was observed in 2003 with the Very Large Telescope
(VLT) Nasmyth Adaptive Optics System NAOS (Rousset et al.~2003) 
attached to the infrared camera CONICA (Lenzen et al.~2003, hereafter NACO) 
as part of the guaranteed time observations of the NACO instrument consortium.
Observations were carried out on 2003 July 22 and 23, 
using the infrared wavefront sensor with the N20C80 dichroic splitting 
the beam such that 20\% of the light serves the adaptive optics control 
loop and 80\% is diverted to the science detector. The infrared-bright star 
Q2 (Glass et al.~1990) with $K=7.3$ mag at a distance of $8{''}$ from the 
cluster center (Q12) served as the natural guide star for the infrared 
wavefront sensor. Images were observed with the 
S27 camera covering a $27{''}\times 27{''}$ field of view with a pixel
scale 27.1 milliarcseconds (mas) per pixel and the $K_s$ filter with 
$\lambda_c = 2.18\mu$m, $\Delta\lambda = 0.35\mu$m, similar to the 
NIRC2 $K'$ filter (Sec.~2.1). Two sets of exposures with different 
integration times were obtained to minimize saturation effects
while ensuring astrometric and photometric performance at the faint end.
Short exposures with detector integration times of 2.0s with 15 co-added
frames aided avoiding saturation of the brighter cluster members, and 20.0s 
exposures with two co-added frames enhanced the sensitivity towards the faintest stars. 
The telescope was moved in 
a wide dither pattern with maximum offsets of $\pm 10{''}$ to avoid artefacts 
due to ghost images of the very bright central cluster stars ($K_s\sim 6$ mag) and to 
facilitate bad pixel correction. Dithering increased the observed field of view to
$40{''} \times 40{''}$, and the central $15{''}\times 15{''}$ of this field 
were extracted for matching with the narrow field NIRC2 data sets. 
A second, shallower epoch of NACO $K_s$ images with 2.0s individual exposure
times and $\pm 7{''}$ maximum dithers 
was obtained in July 2008 with the same instrumental setup under worse 
atmospheric conditions. For all NACO data sets, the FWHM ranged from 
71 to 83 mas with Strehl ratios of 3-12\% (see Table \ref{obstab}).

The raw data were extracted from the ESO archive facility, including 
flat-field and dark calibration frames taken during adjacent nights.
A custom-made pipeline which caters to the special needs of the NACO 
S27 performance was used to reduce the data. The major reduction steps 
are briefly summarized here, while a detailed overview of the reduction 
procedures is presented in Hu{\ss}mann et al.~(2012).
As 50Hz noise is sometimes prevalent in the NACO images\footnote{%
http://www.eso.org/observing/dfo/quality/NACO/
ServiceMode/naco\_noise.html},
manifesting itself in a dense pattern of horizontal stripes moving 
across the images during consecutive exposures, both science and 
sky frames were dark-subtracted and flat-fielded to allow for 50Hz 
correction prior to sky subtraction. A master dark was generated from
3 individual dark exposures with the same exposure time and readout
mode as the science images. The master flat-field was derived from 
sensitivity measurements in each pixel obtained at varying twilight 
flux levels. The master sky was created from sky exposures of semi-empty 
fields observed without AO correction. Residual star flux was removed
from the master sky image by rejecting bright pixels during the combination
of the individual sky frames. As in the case of the NIRC2 data, a bad pixel 
mask was created during the combination of the darks and flat-fields, 
to which individual bad pixel masks are added for each image marking 
cosmic ray events detected with the IRAF {\sl crrej} routine. 
The reduced images are combined using the IRAF/PYRAF task {\sl drizzle}, 
with the individual bad pixel mask applied to 
each frame during image combination. From the 2.0s and 20.0s 2003 and 2.0s 2008 
data sets, one deep image was generated for each by selecting 16, 16, and 33 
individual frames, respectively. The deep images of the 2003 and 2008 2s exposures 
are displayed in Fig.~\ref{quinarch}. The combined sets are termed 2s and 20s 
(combined) images for simplicity in the remainder of the paper.
Auxiliary images were generated from one third of the individual 
science exposures in each of these sets to perform repeated flux measurements 
for a realistic judgement of the astrometric und photometric uncertainties.

\subsection{Astrometry and Photometry}

\subsubsection{Keck/NIRC2}

\begin{figure*}
\includegraphics[width=16cm]{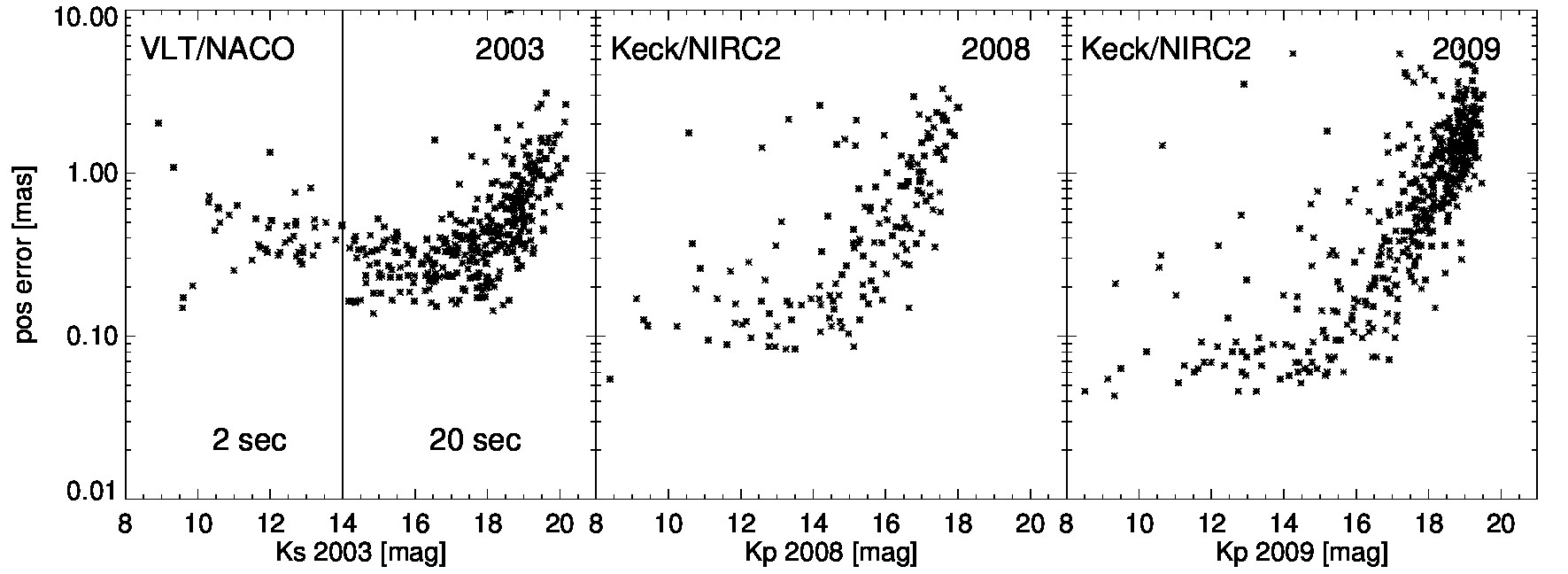}
\includegraphics[width=16cm]{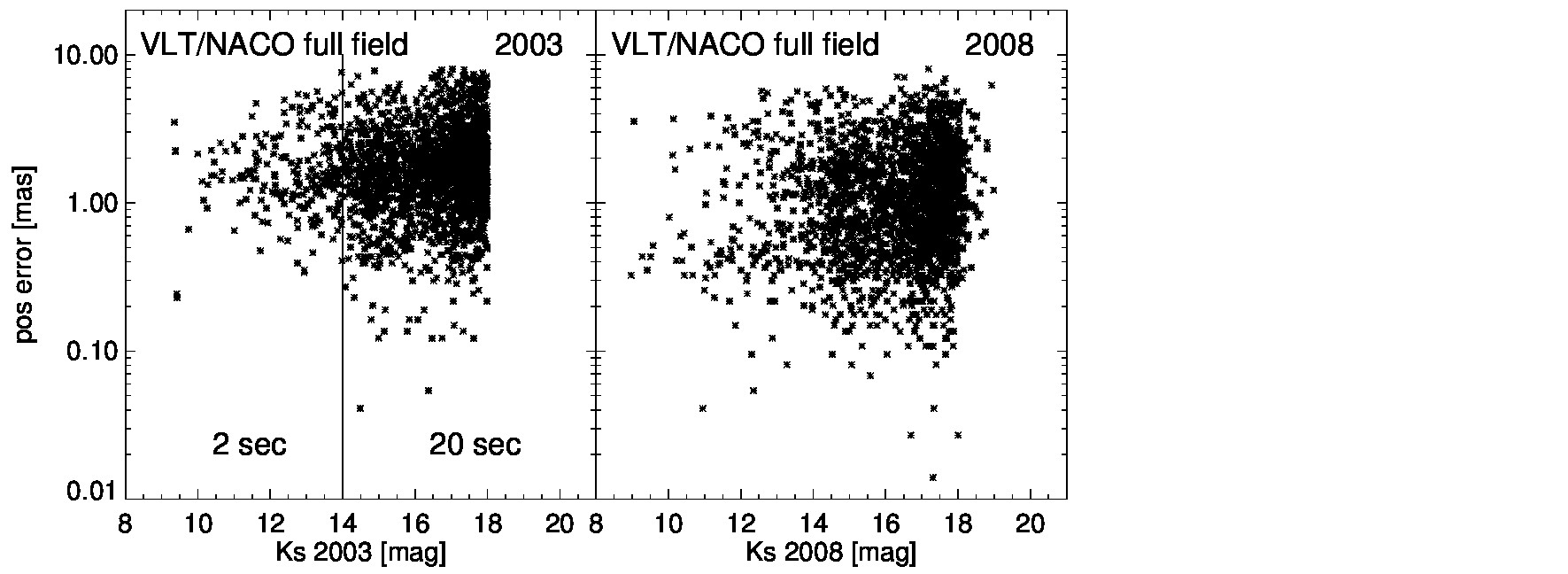}
\caption{\label{poserr} {\sl Top:} Positional uncertainties of the central
VLT/NACO 2003, Keck/NIRC2 2008 and 2009 starfinder astrometry included in the 
proper motion analysis.
{\sl Bottom:} Positional uncertainties of the VLT/NACO full field 2003 and 2008
Ks daophot astrometry with variable PSF across the field. }
\end{figure*}

Positions and fluxes were extracted using the Starfinder crowded field 
point spread function (PSF) fitting tool (Diolaiti et al.~2000). 
In the $10{''}$ NIRC2 field of view, ten isolated sources were 
selected as PSF reference stars. Photometry of all sources
in the field was extracted with a flux threshold of three sigma above the background, 
and three source extraction iterations were performed. In addition to the deep
images, Starfinder was run on each of the three auxiliary images in each epoch. 
As these three images are independent, a repeated 
flux and position measurement is obtained for each star, albeit at the cost of 
photometric sensitivity and astrometric precision due to the fact that each
subset contains only one-third of the images combined in each deep image. 
Stars are required to be detected in at least two auxiliary images in addition 
to the deep science image. For stars detected in all three auxiliary images,
the uncertainty
in the position and magnitude of each star is derived as the root-mean-square (rms) 
deviation from the mean of the three subsets divided by the square root of the 
number of independent measurements, here $\sqrt(3)$, which compensates
for the lack of photometric depth in the auxiliary images as compared to 
the deep science image. In the case that
a star is only detected in two auxiliary images, the uncertainty is derived 
as the deviation from the mean of the two measurements, divided by $\sqrt(2)$.
The positional uncertainties for all data sets are displayed in Fig.~\ref{poserr}.

\subsubsection{VLT/NACO}

\subsubsubsection{Ks 2003 central field}

The larger field of view of $27''\times 27''$ provided by the NACO camera 
shows severe anisoplanatic effects across the field.
These effects were especially pronounced in the 20s images due to the long 
integration time. For the central area overlapping the NIRC2 data, it proved
sufficient to perform PSF fitting photometry with Starfinder in the same 
manner as described above. The extracted field of view covered $15{''} \times 15{''}$,
which allowed for a constant PSF construction with 30 to 40 stars in the 2s and 20s
exposures. The use of the same Starfinder algorithms ensured that
the relative astrometric uncertainty between NACO and NIRC2 was minimized.
Matching the NACO 2003 source list with the NIRC2 2008 and 2009 photometry 
provided a catalogue of 226 sources of which 119 had 3-epoch measurements
suitable for linear-motion fitting (see Sec.~\ref{motionsec}).
This catalogue provides the highest astrometric accuracy available in the Quintuplet
cluster center.

\subsubsubsection{Ks 2003 \& 2008 full field}

For the extraction of the wider field of view with substanially increased source 
counts, the anisoplanatic effects in both the NACO 2003 and 2008 data sets could 
not be ignored. The daophot PSF package with a spatially varying PSF was 
therefore used to perform the photometry on these data sets. While the 2s 
combined images in 2003 and 2008 could be well modelled with a quadratically varying 
PSF, the anisoplanatic effects across the 20s exposures proved neither linear
nor quadratic. To minimize the residual positional errors from anisoplanatism,
the 2003 20s combined image was split into four quadrants and the PSF fitting 
was performed on each quadrant separately. A linearly varying PSF across
each quadrant image provided the lowest flux residuals after PSF subtraction.
The astrometric and photometric source lists of all quadrant images were 
re-combined 
and matched with the 2s 2003
photometry to replace saturated stars with $K_s < 14$ mag in the deep observations.
The combined 2003 $K_s$ source list was then matched with the NACO 2008 daophot photometry
to yield 2134 sources with proper motion uncertainties of less than 3 mas/yr
and $K_s < 18$ mag suitable for fitting the cluster motion.

The three auxiliary images providing the photometric and astrometric uncertainties
were treated in the same way as the deep image, and uncertainties were derived
as in the case of the NIRC2 data discussed above.

\subsubsection{Photometric calibration}

Absolute photometric calibration was obtained by referencing bright stars 
against the UKIDSS source catalogue. 
The UKIDSS survey is a near-infrared $JHK_s$ survey conducted 
with the UKIRT telescope on Mauna Kea, Hawai'i (Lawrence et al.~2007). 
The Galactic center region is covered as part of the Galactic plane survey (GPS)
described in detail in Lucas et al.~(2008). The survey provides a uniform 
spatial resolution of better than 1 arcsecond defined by the median seeing 
conditions at UKIRT (Warren et al.~2007).
The larger NACO field of view is used to identify semi-isolated stars with reliable 
fluxes in the UKIDSS GPS survey. The zeropoint was derived for the combined NACO image
with the short 2.0s individual exposure time to avoid saturation effects. 
Magnitudes reported in the
GPS are derived from aperture fluxes within the FWHM resolution of $1{''}$.
In order to mitigate a systematic error in the zeropoint due to the difference in spatial 
resolution, the flux of stars detected in the NACO frame within the $1{''}$ radius from 
each calibration source was added prior to the zeropoint derivation. This is particularly
crucial as the flux is systematically brighter for UKIDSS stars where the aperture
contains several fainter sources next to the calibration star. As a consequence, the 
uncorrected flux zeropoint is 0.2 mag larger than the zeropoint after correcting for 
the sum of resolved stars in each aperture. To avoid stars partially contributing to 
each $1{''}$ circular aperture, stars located closer than one FWHM (3.2 pixel or 
$0.083{''}$) from the aperture edge were excluded, such that only stars within 
$r_{aper,PSF} = 0.917{''}$ contribute to the total flux of PSF-fitted stars in each
aperture. For most calibration sources, 
corrections due to flux adding are below 0.1 mag, yet the maximum correction reaches 0.3 mag 
in two cases. This constructed aperture flux on the NACO 2.0s image is then compared 
to the UKIDSS photometric catalogue for zeropoint calibration.
The sample of local, reddened stars used as calibrators ensures that color effects are 
minimized, 
and no color terms were found between the Mauna Kea $K'$ and VLT $K_s$ filters.
A total of 15 non-saturated stars with $10.5 < K_s < 12.5$ mag provided a zeropoint of
$23.26 \pm 0.10$ mag.

After calibrating the $K_s$ 2.0s source list to UKIDSS, the 20.0s NACO 
catalogue was referenced to the 2.0s calibrated catalogue using a total of 860 stars
in the common linear regime, $15 < K_s < 17$ mag. The deeper NIRC2 2009 observations
were calibrated with respect to the NACO 2.0s 2003 observations using $\sim 80$ stars in 
the range $11 < Ks < 17$ mag, and the shallower NIRC2 2008 observations were then 
calibrated against the NIRC2 2009 observations with $\sim 130$ stars with
$9 < K^{'} < 18$ mag. The three auxiliary frames of each data set were calibrated
with respect to their respective deep image in all cases. The high number of available
calibration sources between each high-resolution 2003/2008/2009 data set led
to a zeropoint error of less than $\delta K_{ZPT,rms} < 0.01$ mag for each 
calibration, which is consistent with the deviation between all cross-matched data sets 
after zeropointing. The absolute accuracy of the final photometry is therefore limited by 
the calibration with respect to the UKIDSS sample with a zeropoint uncertainty of
$\pm 0.10$ mag.

\subsection{Geometric transformation \& proper motions}
\label{motionsec}

\subsubsection{NACO 2003 \& NIRC2 2008 \& 2009}

Proper motions are derived in the cluster reference frame, adopting our earlier approach as
laid out in Stolte et al.~(2008) and Clarkson et al.~(2012). As there are no known high-resolution
radio sources in the cluster field, we cannot derive the cluster motion in the absolute
reference frame. Systematic effects caused by the necessary choice of the cluster reference
frame, such as the higher number of detected foreground stars as compared to reddened
background sources, are discussed in detail in Sec.~3.2 of Stolte et al.~(2008).

All epochs were transformed
to the NIRC2 2009 image, which served as the reference epoch. The NIRC2 2009 observations
were chosen as astrometric reference because i) the optical distortion solution is 
extremely well known for NIRC2 (Yelda et al.~2010), while it is not sufficiently known 
for the NACO S27 camera, 
and ii) the 2009 observations are photometrically the most sensitive data set,
such that positions have the highest astrometric precision (see Fig.~\ref{poserr}). 
Geometric transformations were calculated individually and iteratively using the IRAF task 
{\sl geomap} for the NACO 2.0s short exposure combined image, the NACO 20.0s deep 
combined image, and the NIRC2 2008 image. An initial estimate of the transformation 
was obtained from bright stars, which are dominated by cluster members.
The proper motion diagram created from this initial guess, 
where cluster stars are concentrated around the origin in the cluster
reference frame, was then used to iteratively improve the transformation.
In the second step, the transformations were refined using all stars with proper motions
within a 2-sigma selection circle around the origin as cluster member candidates.
This selection ensured that only stars not moving significantly with respect 
to the cluster are used to derive the final geometric solution.
Stars with significant motions between the considered epochs are excluded
from the fit.
A second-order polynomial including first-order cross terms provided 
the most accurate geometric transformation solution in all cases. 
The residual rms in the proper 
motion of cluster members after the transformation was applied served as a probe for 
the accuracy of the transformation matrix.  
In the case of the NIRC2 2008 transformation, 38 stars with $11 < K < 16$ mag provided
a residual rms of 0.032 (0.022) pixels or 0.32 (0.22) mas in the x and y directions, 
respectively, in the final geometric solution. For the NACO 2003 2.0s observations,
a final selection of 41 cluster member candidates with $K < 17$ mag yielded a residual rms 
of 0.63 and 0.70 mas in the x and y coordinates, respectively. 
%%% Note: I checked that these residuals are in NIRC2 pixel coordinates
%%% see naco_nirc2_2.0s_ext_k17_poly33_refine.res (same for 15asec soluation, just shifted)
%%% and naco_nirc2_144psfstars_refine_poly33_plot1.ps
%%% Note that 144 stars PSF list is only true for the *full image*, so for the 
%%% ext(racted) image, only 43 PSF stars were used.
For the NACO 20.0s deep combined image, 32 stars with $14 < K_s < 17$ mag led to 
residual rms values of 0.85 and 0.76 mas in x and y, respectively.
As expected, the positional
uncertainty is larger for the lower resolution NACO observations due to the larger
pixel size and the unknown instrumental optical distortions, but is mitigated by the 
longer time baseline. As there is no significant difference between the x and y 
directions, the mean of the x and y rms residuals divided by the time baseline provide
an indication of the proper motion uncertainty from each transformation. 
The contribution of the
geometric transformation to the proper motion uncertainty is 0.27 mas/yr when referencing
the NIRC2 2008 to the NIRC2 2009 positions, and 0.80 mas/5.78 yrs $= 0.14$ mas/yr when 
referencing the NACO 2003 epoch to NIRC2 2009. 
The x and y rms uncertainties of the transformation
are added in quadrature to the individual x and y positional uncertainties to derive
the proper motion uncertainty of each star measured in only two epochs.

For the final proper motion source list, star lists from all epochs were 
matched with the NIRC2 2009 reference list.
For the 5.8 year baseline between NACO and NIRC2, a matching radius of 5 pixels
(50 mas) was used to allow for all field stars to be included in the proper motion 
sample. The NIRC2 2008 positions were matched to the 2009 catalogue with a 
matching radius of 2.0 pixels (20 mas), which accounts for the smaller time 
baseline of only 1 year. 
The final proper motion catalogue contains 119 sources detected in all three 
epochs, and an additional 107 sources only detected in 2003 and 2009.

For the 119 sources detected in all three epochs, proper motions were obtained
from linear fits to the x and y coordinate with respect to the time baseline.
The linear fit was performed with respect to the uncertainty-weighted mean epoch,
%%% this is supposed to be:
%%% total( epochs / sigma_x_or_y^2 ) / total(1/sigma_x_or_y^2)

\begin{center}
$t_{mean} = \frac{\sum(epoch(i) / \sigma_{x/y}^2(i))}{\sum(1/\sigma_{x/y}^2(i))}$, 
\end{center}

\noindent
where $epoch(i)$ is the time, in fractional years, of each measurement at each epoch $i$,
and $\sigma_{x/y}(i)$ denotes the astrometric uncertainty in x or y at the same epoch, 
respectively. 

Note that the dependency of the weighted mean epoch on the x and y positional
uncertainties implies that it is different for each star. The linear fit of
the change in x or y position over time is performed with respect to the
difference between each epoch and the mean epoch, which minimizes
the uncertainty from the intercept and facilitates the derivation of realistic 
fitting uncertainties in the proper motion plane. Otherwise, the 
extrapolation back to zero from an epoch of 2003.56 causes unrealistically 
large uncertainties in the intercept. 

\begin{figure*}
\centering
\includegraphics[width=8cm]{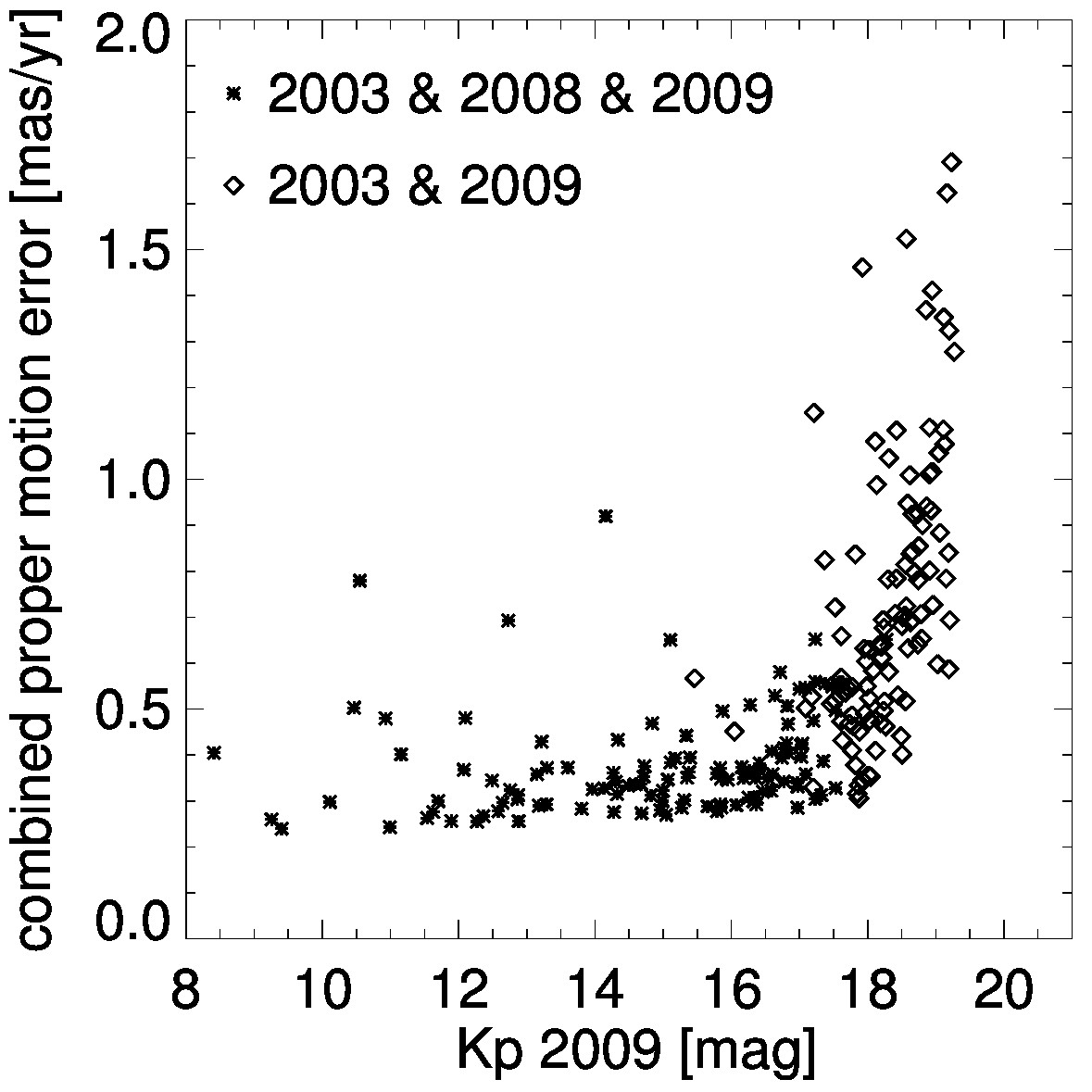}
\includegraphics[width=8cm]{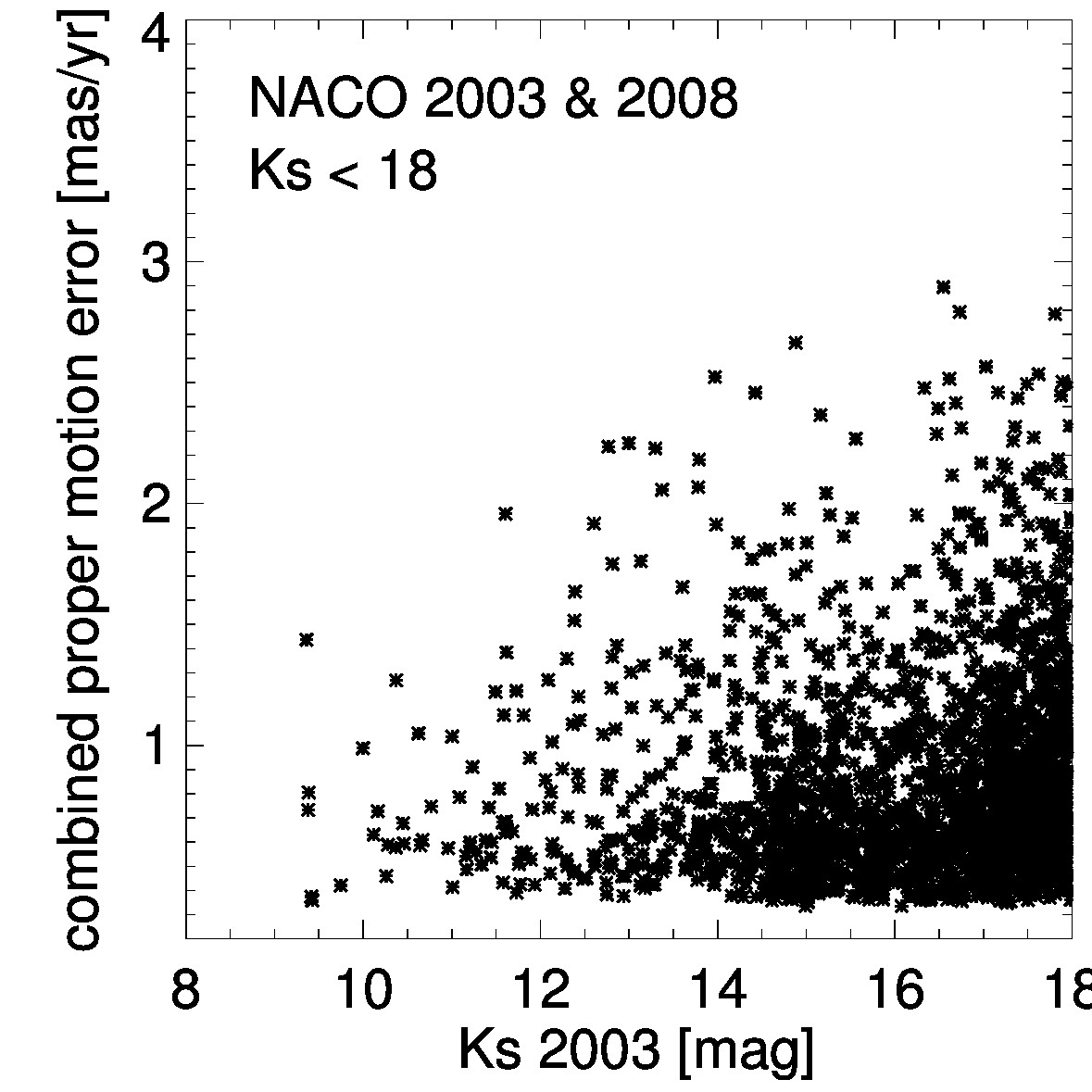}
\caption{\label{pmerrall} 
Proper motion uncertainties of the NACO-NIRC2 and the NACO-NACO 
samples used to derive the cluster motion.
{\sl Left panel:} Combined NACO-NIRC2 proper motion uncertainty of stars matched 
between epochs 2003 and 2009, without detection in 2008 (open diamonds), 
and of stars detected in all three epochs (asterisks). For stars observed
in 3 epochs, the proper motion uncertainty is the uncertainty in the slope 
of the linear motion fit.
{\sl Right panel:} Combined NACO proper motion uncertainty of stars
matched between the 2003 and 2008 epochs. Note the different scales and 
the much larger scatter caused by the lower NACO resolution and the PSF
variation across the full field.}
\end{figure*}

The combined proper motion uncertainties for all data sets are shown 
in Fig.~\ref{pmerrall}.
For stars brighter than $K_s = 16$ mag, the uncertainty is dominated by the 
lower resolution NACO 2003 data set, while between $16 < K_s < 17$ mag, the
shallower NIRC2 2008 data determine the proper motion uncertainty. 
Because of the significant difference in detection sensitivity, stars with $K_s > 17$ mag
are only detected in the 2003 and 2009 epochs. The combined proper motion uncertainty
of these sources is shown as open diamonds in Fig.~\ref{pmerrall} (left panel).

As the fainter field stars were predominantly lost in the 2008 observations,
the 3-epoch sample is heavily biased towards cluster stars. In order to measure
the relative motion between the cluster and the field, we therefore had to 
include the faint field stars detected in the 2003 and 2009 epochs alone.
For the 107 sources not detected in 2008, the proper motion had to be 
estimated from the positional difference between 2009 and 2003, and no 
goodness-of-fit evaluation was possible. As most of these stars are fainter 
than $K_s = 17.5$ mag, the larger positional uncertainties are reflected 
in larger proper motion uncertainties than in the case of the 3-epoch 
linear-fitting proper motions. 

The proper motion uncertainty for stars measured in all three epochs is determined from 
the linear fitting error of the slope of the three measurements, leading to a 
median uncertainty of 0.34 mas/yr for stars with $K_s < 17$ mag 
for the 3-epoch sample (Fig.~\ref{pmerrall}, left panel, asterisks).
The median proper motion uncertainty for stars detected in 2003 and 2009, 
but not in 2008, was 0.66 mas/yr (Fig.~\ref{pmerrall}, left panel, open symbols).

\subsubsection{NACO 2003 \& 2008 full field}

The same iterative procedure is employed when matching the NACO 2003 and 2008 
epochs covering the full $40{''}$ field of view. 
For these two data sets, obtained with the same CONICA S27 camera
and hence similar optical properties, the geometric transformation resulted in
a residual x and y rms of 1.1 and 1.3 mas. These transformation uncertainties
contribute 0.24 mas/yr to the final astrometric uncertainty, which is 
dominated by the positional uncertainties of the PSF fitting procedure in 
both NACO epochs (see Fig.~\ref{poserr}, bottom two panels). 
After matching the NACO 2003 \& 2008 source catalogues, the
median proper motion uncertainty of stars with $K_s < 18$ mag in both x and y 
results in 0.5 mas/yr for the time baseline of 5.0 years (Fig.~\ref{pmerrall}, 
right panel). The final proper motion catalogue contains
2137 sources across a combined field of view of $41{''} \times 41{''}$.

\subsection{Combined proper motion catalogues}

The source counts of all proper motion catalogues are summarised in Table \ref{pmtab}.
In the final proper motion source list as published in Table \ref{nirc2tab}, 
the central cluster area is covered with the NACO-NIRC2 astrometry and photometry of
226 sources covered in 2003, (2008), and 2009, providing
the highest astometric performance. All sources in this catalogue have proper motion 
uncertainties of less than 1.5 mas/yr. In addition, a more complete coverage of 
the cluster center is provided due to the sensitivity of the NACO-NACO sample
in Table \ref{nacotab}, albeit at the cost of astrometric accuracy.
The outer cluster areas contain NACO-NACO astrometry exclusively, and the
full NACO-NACO catalogue contains 2137 stars with proper motion 
uncertainties of less than 3 mas/yr and $K_s < 18$ mag. 
Sources with larger uncertainties are not included in the final NACO-NACO catalogue. 
The full version of both tables is available electronically from the 
cds database\footnote{Full data tables are available from the journal webpage.}.

\section{Proper motion analysis}
\label{pmsec}

In this section we derive the orbital motion of the Quintuplet cluster. 
We will then use the knowledge of the 3D velocity of the cluster to constrain 
the cluster orbit in the central Galactic potential.
An upper limit to the internal velocity dispersion is also provided.

\subsection{Quintuplet's orbital motion}
\label{orbmovesec}

\begin{figure*}
\includegraphics[width=8.4cm]{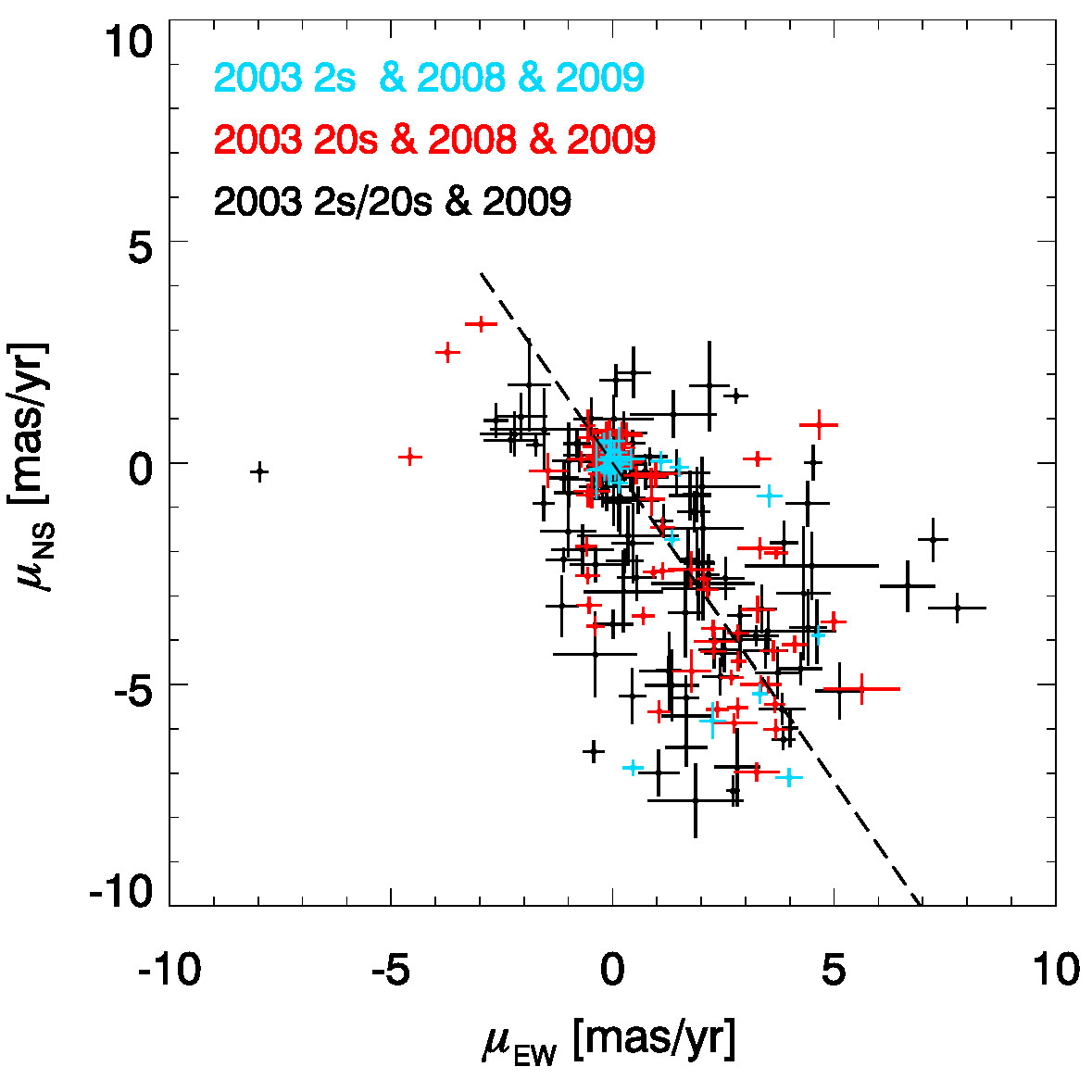}
\hspace*{4mm}
\includegraphics[width=8.4cm]{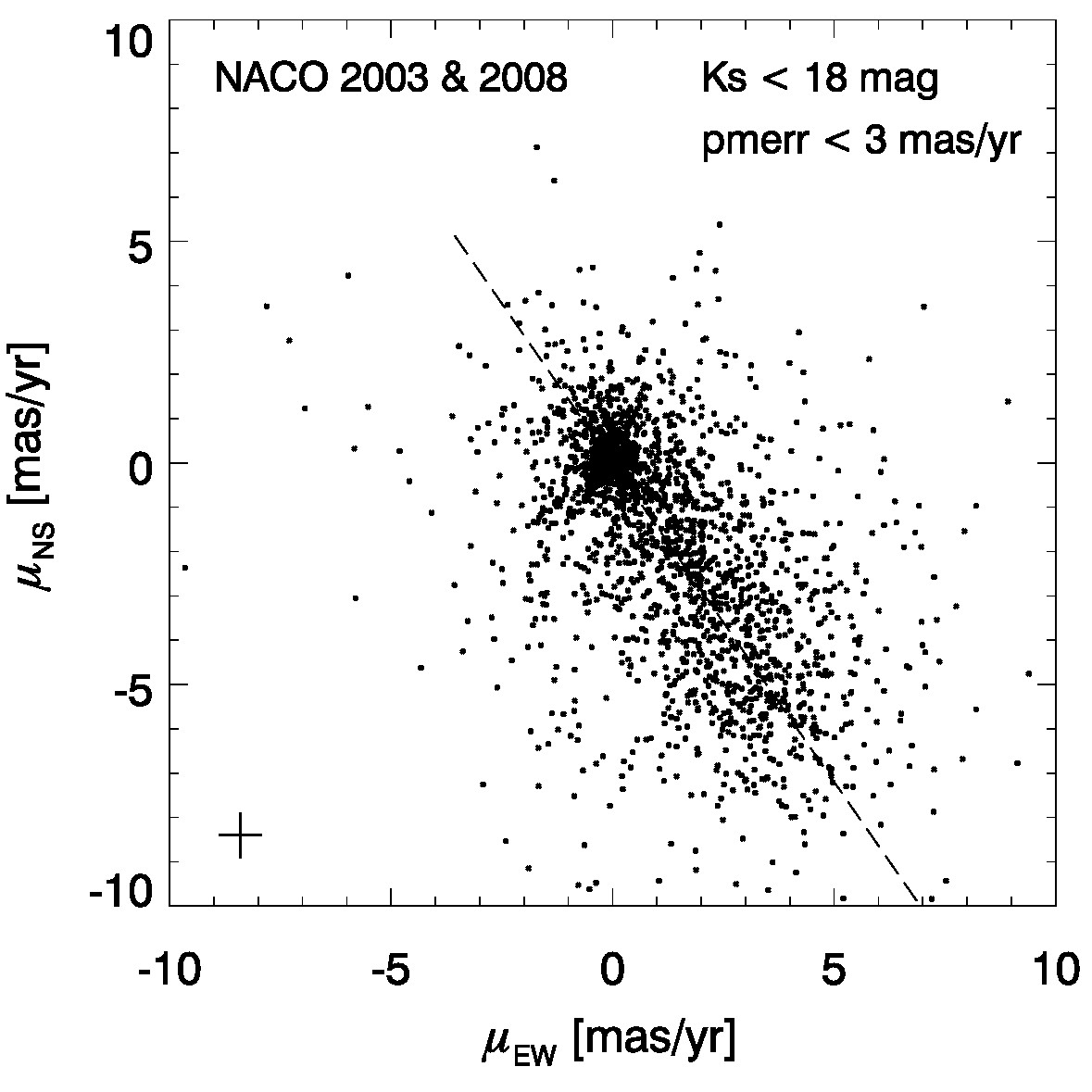}
\caption{\label{pmdia} 
{\sl Left:} Proper motion diagram of all 226 sources in the common $10{''}$ 
NACO-NIRC2 field of view. Sources with measurements in all three epochs dominate in the 
concentrated clustering around (0,0), which represents the best sample of cluster 
member candidates. The dashed line corresponds to the orientation of the Galactic 
plane. Field stars are on average significantly fainter, and frequently
only detected in the deeper 2003 and 2009 observations, as evidenced in their larger
motion uncertainties. 
{\sl Right:} Proper motion diagram of all 2137 sources detected in the full NACO
$41{''}$ combined field of view. Only stars contributing to the cluster motion fit 
with $K_s < 18$ mag and $\sigma_\mu < 3$ mas/yr are shown. The cross in the lower
left corner depicts the median proper motion uncertainty.}
\end{figure*}

The proper motion diagram with all 226 central sources is shown in Fig.~\ref{pmdia}
(left panel),
and the proper motion distribution of all stars in the extended NACO field is 
shown in Fig.~\ref{pmdia} (right panel).
As cluster members are on average brighter, they dominate the dense clump of 
stars around the origin, which implies zero motion with respect to the cluster 
reference frame as defined above. Hence, these stars are cluster member candidates, 
denoted cluster members for simplicity in the following. The absolute motion 
of the cluster with respect to the field is obtained for the two proper motion 
samples individually. The NIRC2-NACO sample provides the more accurate astrometric
measurements while being limited to a small number of field reference stars.
The NACO-NACO sample, on the other hand, contains a ten times larger number of 
stars while being limited by the larger proper motion uncertainties.
In the case of the field stars used to derive the relative 
motion between the cluster and the field population in the inner bulge, 
the uncertainty in each individual measurement does not influence the 
fit to the ensemble substantially. The velocity dispersion, however, is 
derived from the standard deviation of cluster members, and is therefore
strongly influenced by the individual motion uncertainties.
It is crucial for the 
derivation of an upper limit to the velocity dispersion that the sample with the
smallest proper motion uncertainties be used, such that the motions are
reliably determined. We therefore employ the NACO-NIRC2 sample of the central 
cluster to derive both the absolute motion of the Quintuplet with respect to 
the field and to constrain the velocity dispersion. From the NACO-NACO sample 
of the extended field, an independent estimate of the absolute cluster motion 
is obtained using a large sample of field reference stars.

\subsubsection{Fitting the cluster proper motion}

Following the procedures developed in our previous investigation of the Arches 
cluster (Clarkson et al.~2012), we have employed a binning-independent fitting 
method to the proper motion distribution using Expectation Maximization (EM). 
% with gaussian priors. 
As a statistical method, EM is particularly useful for sparse or incomplete datasets.
This is particularly the case for the low number of field stars as compared to 
cluster stars in the NIRC2-NACO sample.  
This method allows us to derive the probability of a star occupying a given 
location in proper motion space to belong either to the cluster or to the 
field distribution (see Clarkson et al.~2012, equ.~1). 
Taking into account the proper motion uncertainty of each star, 
individual membership likelihoods are also derived. 
The distribution of sources in the proper motion diagram is modelled by 
two bivariate gaussian functions, and the best-fit 
model is derived from EM fitting following Bishop (2006, Chapter 9,
see also Press et al.~2007, Chapter 16, pp. 842). 

Two elliptical gaussian functions are fitted to the ensemble of field and 
cluster stars in the proper motion plane, 
with one gaussian representing the cluster and one the field. Cluster and field 
distributions are allowed to overlap. No further constraints need be assumed 
{\sl a priori} for the fit. Even with relatively large deviations 
of the initial guess from the final solution,
the two gaussians converge towards the same cluster and field solution.
The peak distance between the two elliptical gaussians yields the absolute motion 
between the cluster and the field sample, while
the semi-major axis of the cluster ellipse provides an estimate of, 
or an upper limit on, the internal velocity dispersion.
The fitting method is described in numerical detail in Clarkson et al.~(2012)
and follows the procedures established for the Quintuplet cluster in 
Hu{\ss}mann (2014), and only the major features are recaptured here.
\\

In addition, the membership probability of each star is derived taking into 
account the individual uncertainties in the proper motion of each star
following the procedures in Kozhurina-Platais et al.~(1995).
In the minimization procedure to fit the relative motion of the Quintuplet
with respect to the field,
all stars are included in the fit of the cluster and field motion ellipses,
and no distinction is made between cluster and field stars. Hence, for the 
derivation of the motion the membership probabilities are not relevant.
Nevertheless, these membership indicators are included in Tables \ref{nirc2tab} 
and \ref{nacotab} for future reference. 
\\

% Results of Benjamin's motion fits - both for Keck & NACO F1 :)

\begin{figure*}
\hspace*{-1cm}
\includegraphics[width=9.4cm]{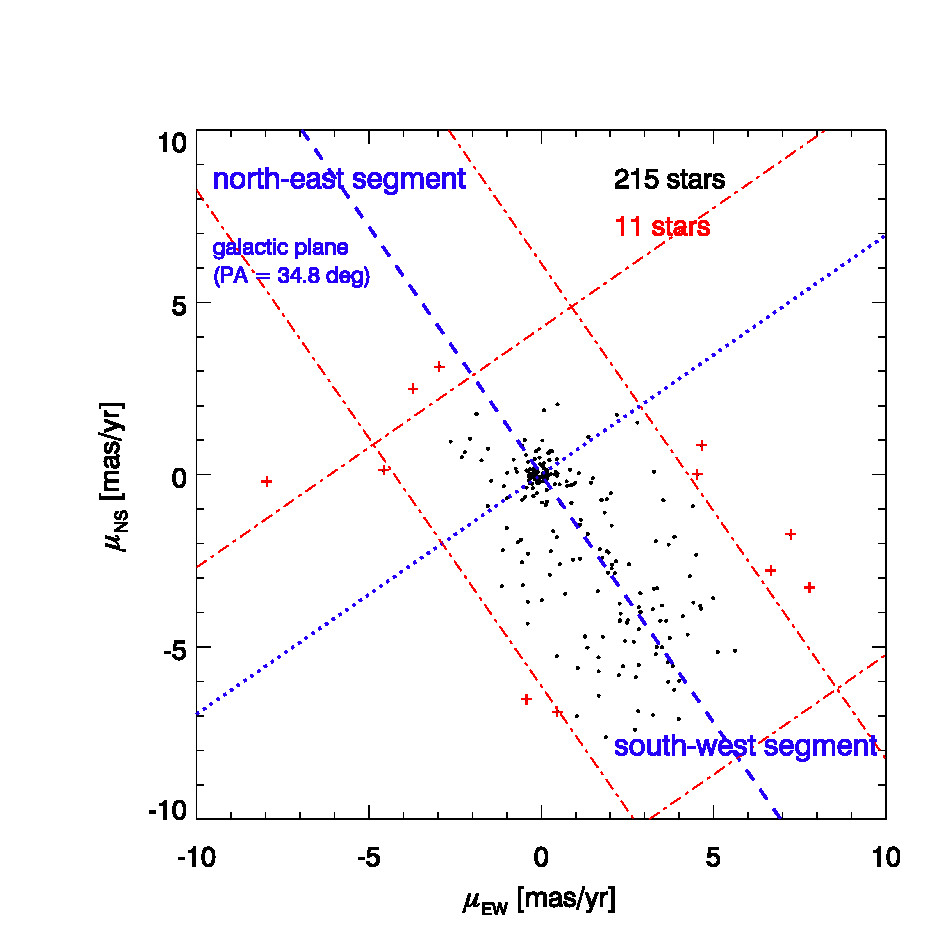}
\includegraphics[width=9.4cm]{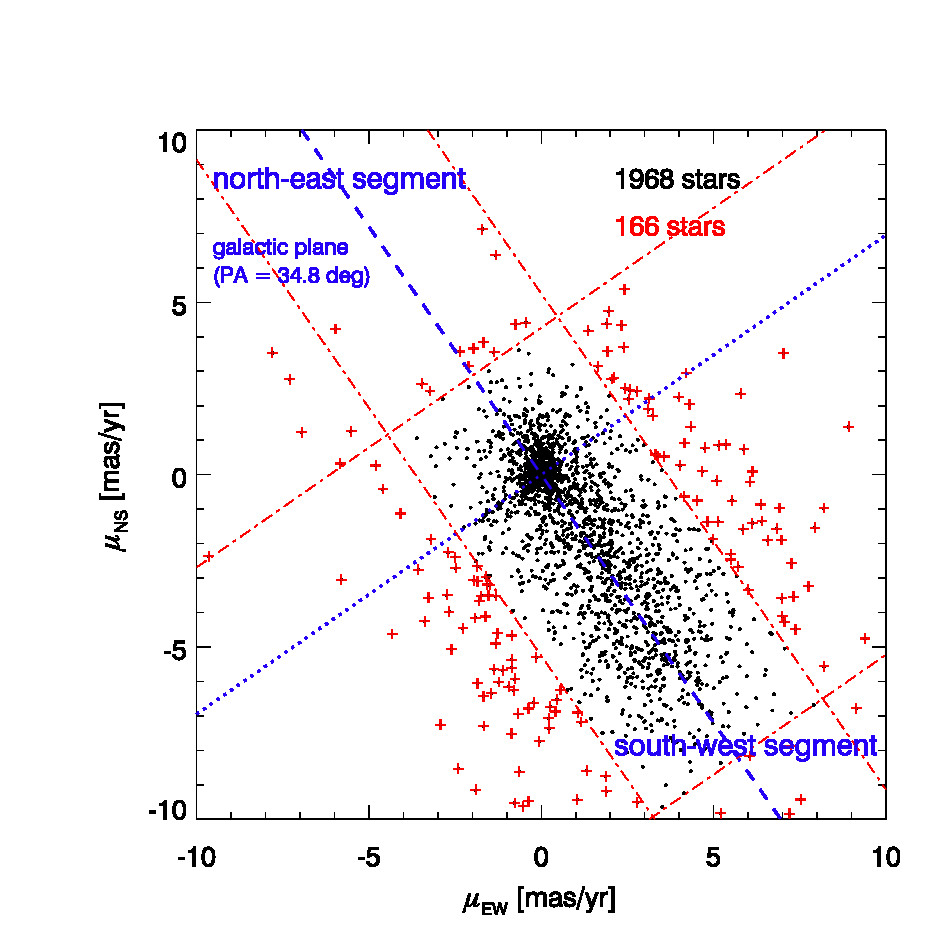}
\caption{\label{moveselect}
Proper motion diagrams of the Quintuplet cluster. 
{\sl Left panel}: The highest-resolution Keck observations of the cluster core
($r < 0.2$ pc). The dashed line depicts the orientation of the Galactic plane, 
and the red dash-dotted lines display the selection criteria
for stars to be included in the cluster motion fit. The fact that 
the extended tail of field stars is distributed along the Galactic
plane indicates that the cluster moves along the plane.
{\sl Right panel}: Proper motion diagram derived across the full 1 pc field
of the central Quintuplet cluster. The lower resolution and hence astrometric
performance of the NACO
observations are evidenced in the larger scatter both in the extended
cluster profile as well as in the dispersed field population. The large
number of field and cluster stars mitigates this disadvantage when fitting
the cluster motion.}
\end{figure*}

\begin{figure*}
\vspace*{-3.4cm}
\hspace*{-0cm}
\includegraphics[width=8cm]{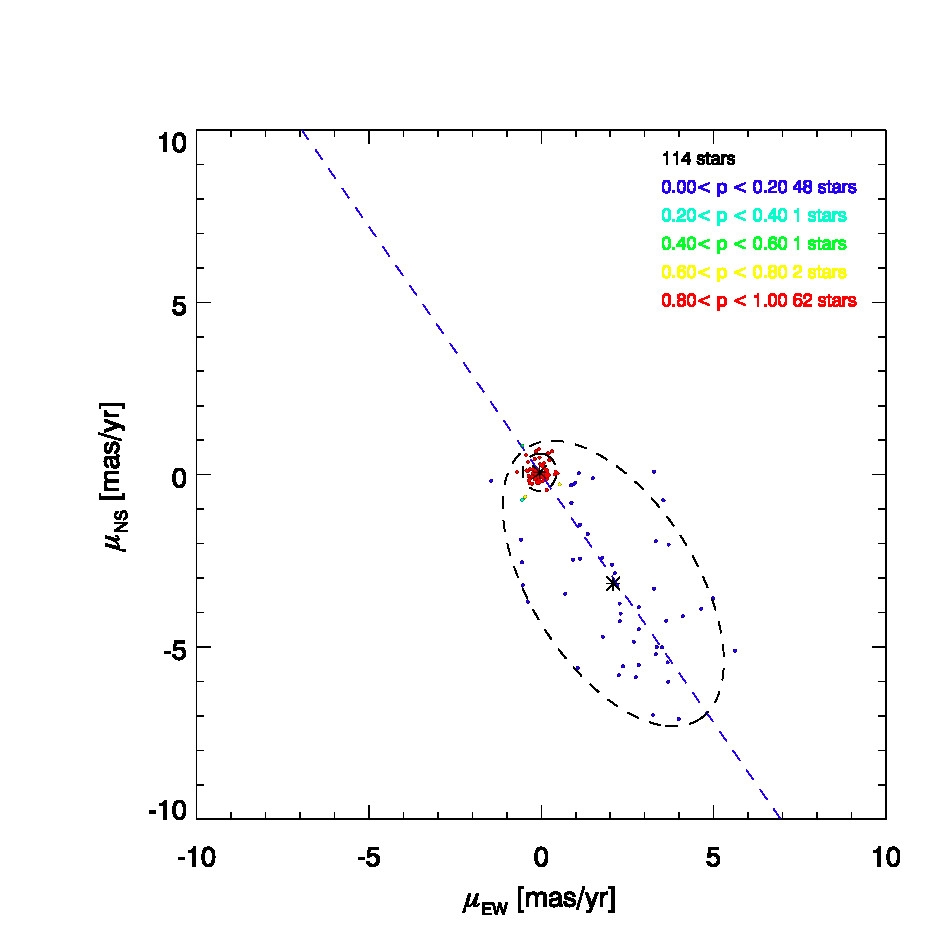}
\includegraphics[width=8cm]{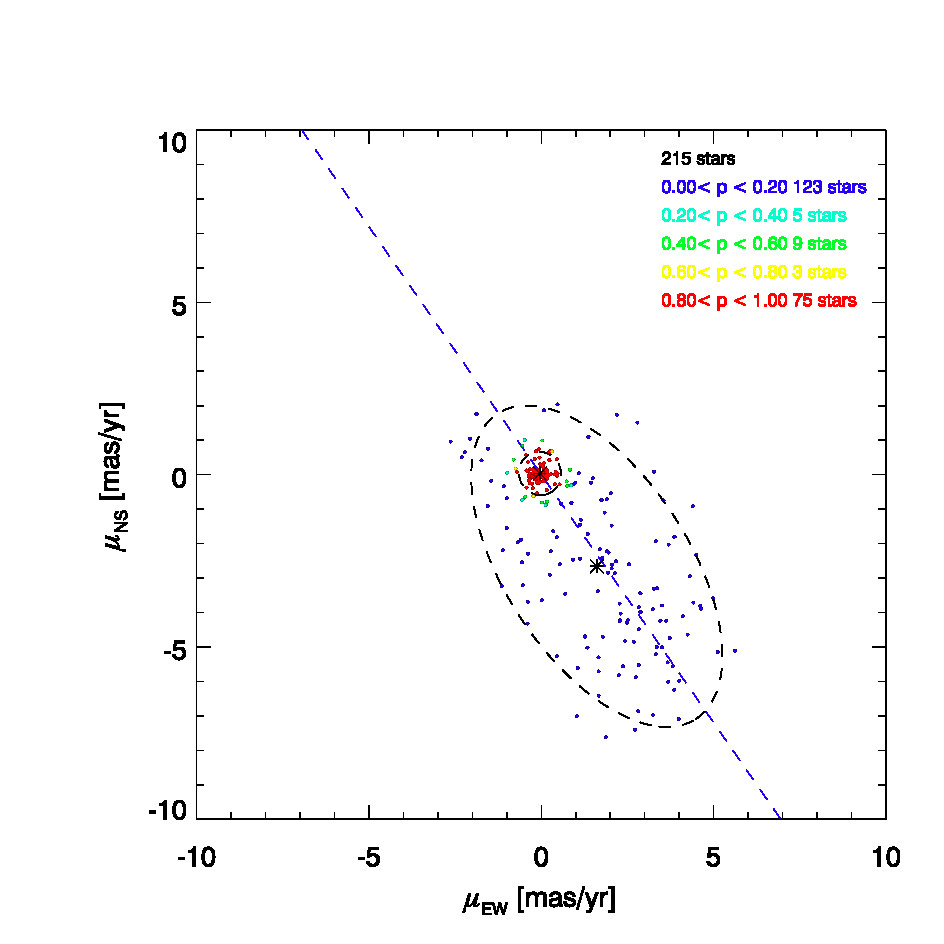} \\
\includegraphics[width=8cm]{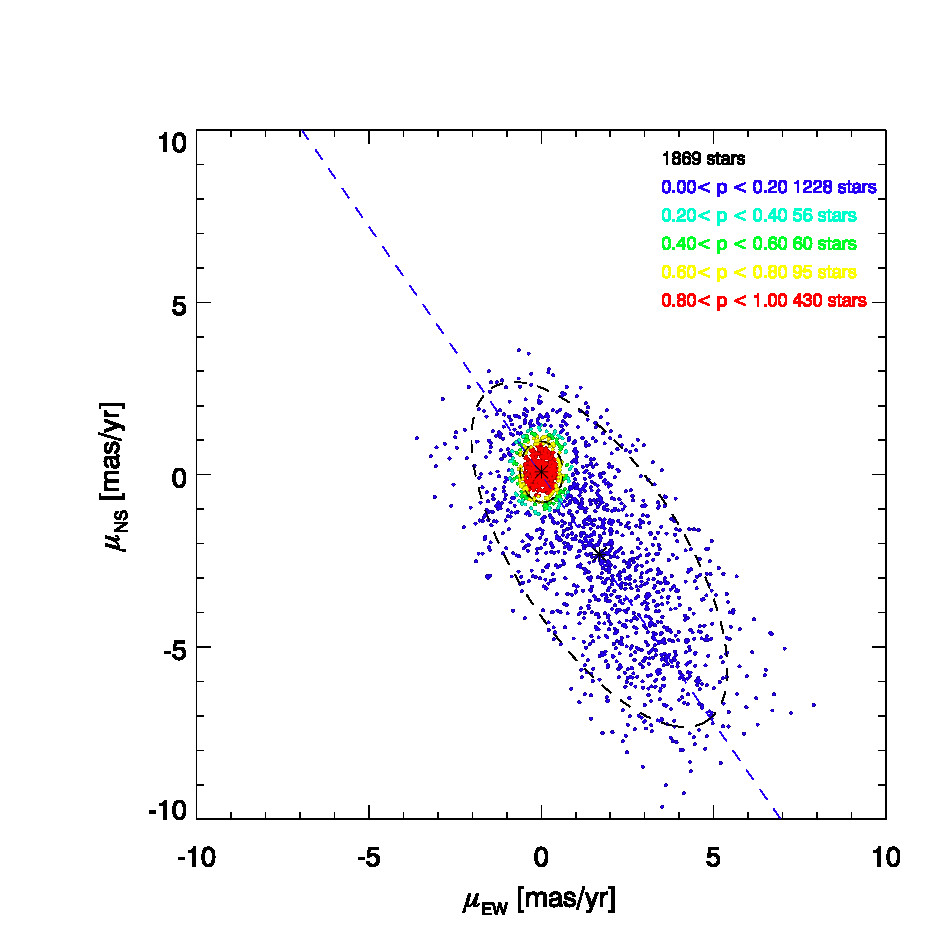}
\includegraphics[width=8cm]{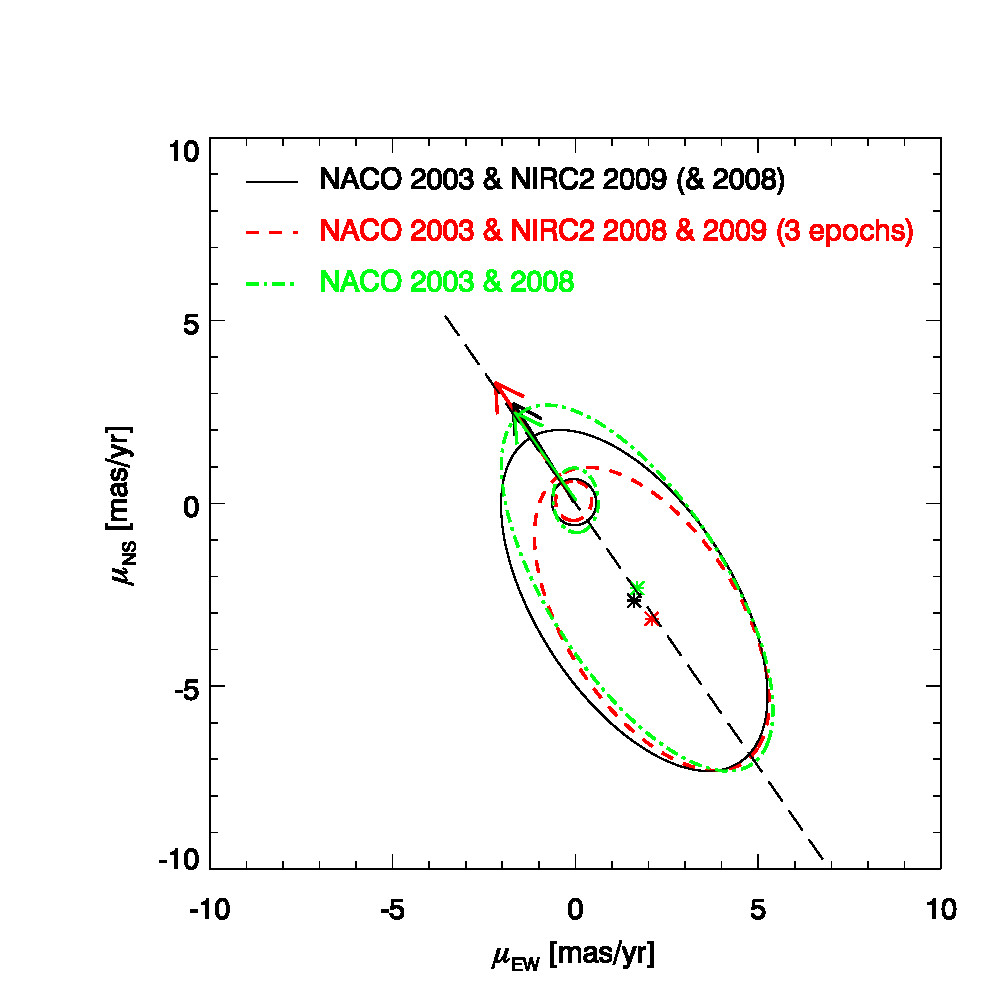}
\caption{\label{movefit}
Proper motion diagrams of the three fitting samples including the
two gaussian fits to the cluster and the field distributions.
All fitting parameters are provided in Table \ref{fittab}.
The location of the cluster's motion along the major axis of the field 
ellipsoid confirms the motion of the Quintuplet along the Galactic plane 
(dashed line). The colour coding represents membership probabilities.
The fact that the red sample is highly concentrated at the origin reveals
the excellent astrometric performance in the high-resolution Keck observations
(upper panels). 
In all panels, the centroid distance of the field and cluster ellipsoids 
(black asterisks) yields the measurement of the bulk motion of the cluster
with respect to the field. 
{\sl Top left:} The bivariate gaussian fits of the
3-epoch sample only. Note the low number of field stars available in this 
sample. {\sl Top right:} All astrometric sources with 2 or 3 epoch measurements
in the NIRC2-NACO proper motion catalogue. The number of faint field stars
is significantly enhanced, rendering the relative motion between cluster and 
field more reliable. 
{\sl Bottom left:} The larger uncertainties in the NACO-NACO proper motions 
cause the cluster ellipsoid to be elongated, where the orientation 
of the ellipsoid is not physically significant, as expected. 
{\sl Bottom right:} The three fits to the cluster and field ellipses 
are overlaid to illustrate the absolute uncertainties.
The arrows indicate the proper motion of the Quintuplet with respect to the field.}
\end{figure*}

As outliers tend to skew the elliptical fit, especially to the 
field distribution, stars significantly distant from the Galactic 
plane were removed from the fit. 
Stars were rejected if their proper motion vertical to the Galactic 
Plane was larger than $\pm 3.5$ mas/yr, or if their motion parallel to the 
plane was larger than $+3.5$ mas/yr or more negative than $-10$ mas/yr to 
remove outliers
from the field sample (Fig.~\ref{moveselect}, see also Hu{\ss}mann 2014
for a detailed explanation). Only stars with proper motion 
uncertainties of less than 1.5 mas/yr are included in the NIRC2-NACO 
source list (see Sec.~\ref{motionsec}), such that the scattered distribution
of stars beyond the selection limits does not originate from particularly large
proper motion uncertainties in these objects. Therefore,
these stars are likely rapidly moving foreground interlopers. 
All remaining stars were then fitted with a field and cluster ellipsoid 
in the shape of two bivariate gaussian functions simultaneously,
where both minor and major axes parameters and centroids as well as position
angles are unconstrained. 
In Fig.~\ref{movefit}, we show the two bivariate gaussian fits to the 
proper motion diagram, and fitting parameters are given in Table \ref{fittab}. 
Likely cluster members are shown in red.  
The Quintuplet's proper motion is measured as the distance between the
centroids of the field and the cluster ellipsoids.
Fitting all remaining 215 stars in the NIRC2-NACO sample yields a bulk motion 
of $\mu = 3.16$ mas/yr for the Quintuplet cluster with respect to the field, 
which corresponds to 120 km/s at the GC distance of 8.0 kpc. 
Including only the 119 stars measured in 3 epochs, and hence 
the most accurate proper motion ensemble from linear motion fitting, 
results in a bulk motion of $\mu = 3.87$ mas/yr or 147 km/s. 
The large difference between the two values reflects
the sensitivity of the bulk motion to the centroid of the extended
field ellipse, which is particularly sensitive to changes in the 
distribution of field stars in the proper motion plane and hence the 
sample selection. The left panel of Fig.~\ref{movefit} illustrates 
the sparse field population contributing to the centroiding distance
between the cluster and the field ellipse. The 3-epoch sample yields the
maximum proper motion of the cluster along the Galactic plane, and hence
suggests this value is an upper limit to the true 1D motion of the Quintuplet.
The position angle of the field ellipse is fitted
to be 32 to 33 degrees in all samples, which is in excellent agreement with the 
position angle of 34.8 degrees of the Galactic plane. 
The proper motion of the Quintuplet is therefore consistent with
a cluster orbit oriented along the Galactic plane, with no evidence for a 
significant motion component out of the Galactic plane.

%%% use motions from vl vb fit here:
%
% Keck (all stars in sample, see keck fitting values): 
% PA field = 32.96 vs. expected 34.7 (following Benjamin's thesis)
%  => delta = 1.7 deg   &&  v_2D = 119.78 km/s  (= "separation")
% formally: vl = cos(delta) * v_2D = 119.727 km/s
%           vb = sin(delta) * v_2D =   3.55 km/s 
%
% for completeness and reference I'll leave this here:
%The angle between the orbital motion vector and the Galactic plane is
%provided by the fit to be $32.96\degree$, which is deviating
%from the exptected orientation of the Galactic plane ($34.7\degree$) 
%by only $\delta = 1.7\degree$. This would lead to formal motions of
%$v_l = 119.7$ km/s along the Galactic plane and $v_b = 3.5$ km/s
%perpendicular to the plane, where the latter value is significantly lower
%than the uncertainty associated with the motion fit (see below).
%We conclude that  
%the Quintuplet cluster moves almost exactly along the Galactic plane.

Given the sensitivity of the fit to the sample selection, we have used 
the wider, albeit less well resolved, NACO field coverage to verify the 
derived velocity of the cluster with respect to the field. 
The same proper motion selection was applied as 
above (Fig.~\ref{moveselect}, right panel), 
as less restrictive selection criteria did not influence the 
fit. 
Only stars with proper motion 
uncertainties of $\sigma_\mu < 3$ mas/yr were included in the fit.
From the 1968 stars with $K_s < 18.0$ mag selected in the central 1 pc square 
Quintuplet field, a bulk motion of 3.08 mas/yr or 117 km/s is found,
and the motion is again with a position angle of 33 degrees aligned 
along the Galactic plane. The larger proper motion uncertainties 
intrinsic to the NACO observations are reflected in the 
more extended cluster member distribution around the origin as compared
to the member candidates in the NIRC2-NACO sample (upper vs.~lower panel
in Fig.~\ref{movefit}). Correspondingly, a larger number of cluster stars 
scatter into the field population.
The centroid of the field ellipse is therefore pulled 
towards the center of the cluster ellipse (i.e.~the origin of the 
proper motion plane, Fig.~\ref{movefit}), such that the absolute motion 
of the cluster with respect to the field is estimated to be smaller than in 
the case of the Keck data. This value can hence be considered a lower limit
to the cluster motion along the Galactic plane. 

\begin{figure*}
\center
\includegraphics[width=12cm]{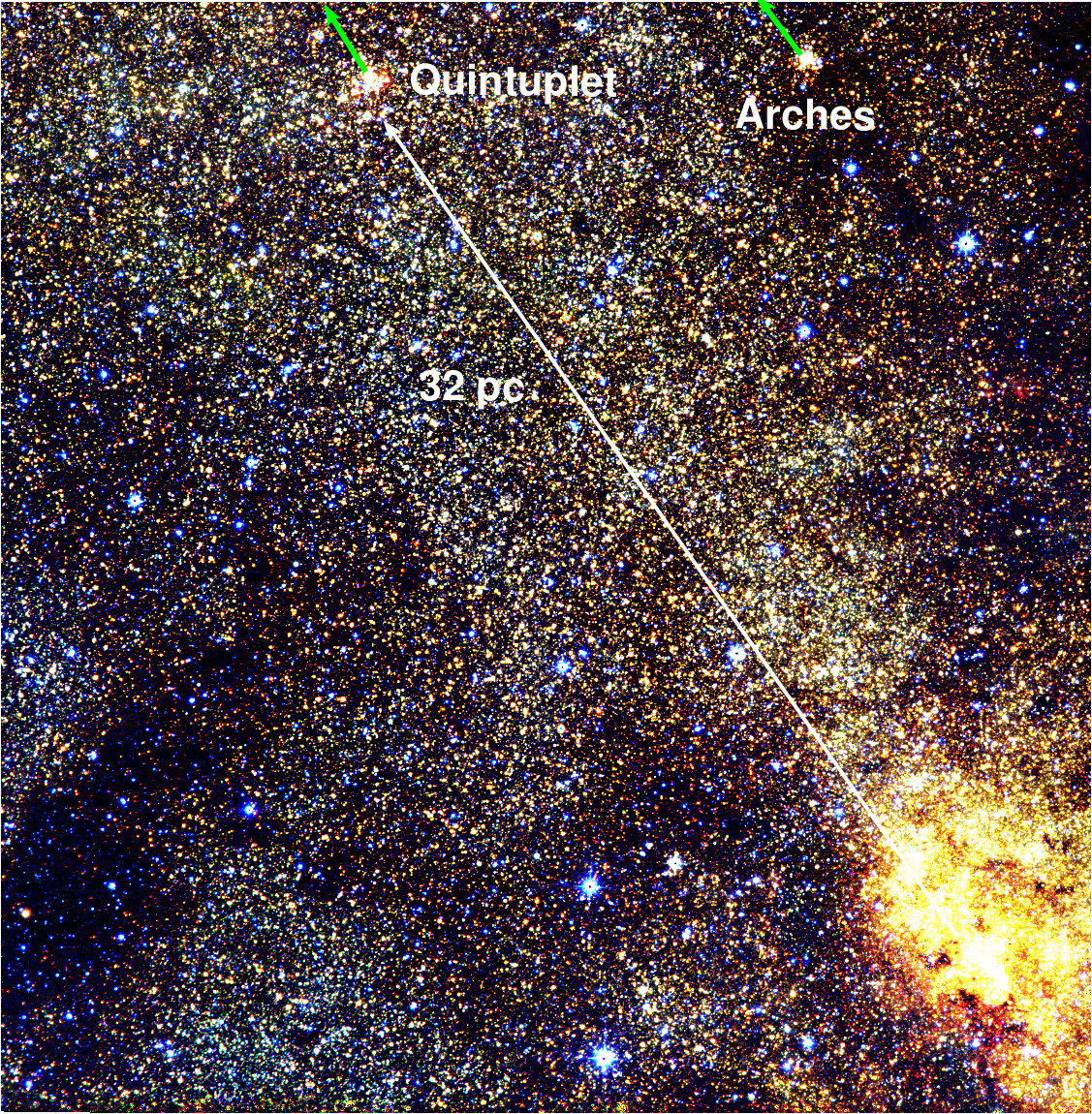}
\caption{\label{archquin_jhk}
UKIDSS $JHK$ colour composite of the Galactic center region. The motions of the 
Arches and Quintuplet clusters parallel to the Galactic plane are indicated by 
the arrows. While the Quintuplet is located almost on the Galactic disk,
the Arches is located at a projected distance of 10 pc above the disk. The infrared-bright 
area at the bottom (South-West) of the image is the nuclear cluster. North is up, and 
East is to the left.}
\end{figure*}

Combining all three fitted values provides
an absolute uncertainty to the orbital motion of the Quintuplet cluster.
We therefore conclude that the 2D motion of the Quintuplet cluster, 
i.e.~the relative motion of stars in the cluster reference
frame with respect to the surrounding field population, 
is found to be $132 \pm 15$ km/s along the Galactic plane.
This 3D velocity is similar to the orbital velocity of the 
Arches cluster and is oriented along the Galactic plane, 
as illustrated in Figure \ref{archquin_jhk}.

\subsubsection{Quintuplet's 3D orbital motion}

In order to derive the 3D orbital motion of the Quintuplet cluster with 
respect to the field, we assume that the field is on average at rest.
As we discussed in Stolte et al.~(2008), the relative motion between 
blue and red fore- and background stars in the Arches field sample
was found to be in excellent agreement with the velocity deviation 
observed in bulge giants (Sumi et al.~2003, see especially the red clump 
sample in their Fig.~8). 
This consistency implies that the field population consists of a 
representative sample of bulge motions
along the line of sight, which are on average at zero velocity with 
respect to the Sun. We therefore concluded that the mean motion 
of the detected field stars is consistent with the field being at rest.
As the Quintuplet field sample is very similar in velocity space and in 
number to the Arches field sample, as expected from the identical 
observational setup, the field sample is also considered to be 
at rest, with the cluster moving with respect to this field.
A detailed discussion of the field contribution in the center 
of the Quintuplet cluster is provided in Hu{\ss}mann et al.~(2012). 
The CMD of field stars is entirely dominated by red bulge giants
at the faint end, $H > 19$ mag, and red clump stars at intermediate
magnitudes, $16 < H < 18$ mag (see their Fig.~8, left panel).
Only very few Galactic disk sources, clearly discerned due to their blue
colors at $H-K_s < 1.3$ mag, contaminate the field sample. 
These Galactic disk sources are expected to be on the flat part of the 
Galactic rotation curve and tend to have comoving orbital velocities 
of $v_{3D,circ} \sim 230$ km/s, on the same order as the clusters with respect 
to the bulge population. We therefore assume that the centroid 
of the reference velocity ellipsoid is not biased by disk stars.
As they comprise a minor fraction of field stars of at most 
a few percent, they do not influence the derivation of the 
relative motion between the cluster and the field. 
With the assumption that the field reference sample is on average at rest,
the mean apparent proper motion of the field in the proper motion 
plane represents the absolute 2D motion of the cluster through
the bulge.

The radial velocities of stars in a wider Quintuplet field were reported 
by Liermann et al.~(2009). In order to deduce the mean radial velocity of the 
young cluster population, we include only the early-type cluster members
from their sample and we exclude the Wolf-Rayet stars and other stars 
with uncertain radial velocity measurements.
From this sample with reliable velocity measurements, the mean radial velocity 
of the 52 early-type cluster members is found to be $102 \pm 2$ km/s 
(the standard deviation of 13 km/s is divided by sqrt(52)
to obtain the standard error of the mean, whic hwe use as the radial
velocity uncertainty). 
Note that the standard deviation of the radial velocity measurements is 
dominated by the fitting accuracy to the line centroids in the spectral fits 
(see Liermann et al.~2009 for details), and does not provide an independent
estimate of the (radial) velocity dispersion of the cluster.
Combining the proper motion of $132 \pm 15$ km/s with this radial velocity, 
we derive the present-day 3D orbital velocity of the Quintuplet cluster 
to be $167 \pm 15$ km/s. 
This 3D velocity is similar to the orbital velocity of the Arches cluster 
and is oriented along the Galactic plane, as illustrated in Figure \ref{archquin_jhk}.

\subsection{Quintuplet's velocity dispersion}
\label{veldispsec}

%%% include: mean xpm/ypmerr uncertainties are on the same order
%%% as the measurements: mean(xpmerr(65 clus stars)) = 0.2592 mas/yr
%%%                      mean(ypmerr(65 clus stars)) = 0.2527 mas/yr
%%%
%%% medians are slightly lower, but this suggests it's still not a 
%%% a real measurement (medi(xpmerr) = 0.243, medi(ypmerr) = 0.229 mas/yr)
%%%

As discussed above, the most accurate proper motions least affected 
by residual astrometric uncertainties are given by the 3-epoch sample.
Of the 119 stars in this sample, 55\%, or 65 stars, are found to be likely 
proper motion members (red points in Fig.~\ref{movefit}). For these cluster
candidates, the velocity dispersions in the x- and y-direction are measured
to be $\sigma_x = 0.271$ mas/yr and $\sigma_y = 0.246$ mas/yr, corresponding
to 10.3 km/s and 9.3 km/s, respectively. The mean proper motion uncertainties
for this sample of cluster members are with
$xpm_{err,mean} = 0.259$ mas/yr and $ypm_{err,mean} = 0.253$ mas/yr, about the
same as the fitted dispersions, suggesting that the dispersion measurement 
is dominated by the individual motion uncertainties. The values therefore
comprise an upper limit to the intrinsic velocity dispersion of the Quintuplet 
cluster. 

% revision 10.4.2014
The fact that the mean uncertainty of 0.253 mas/yr in the y-direction 
is larger than the apparent velocity dispersion measured in cluster members
suggests that the individual uncertainties are slightly overestimated.
The uncertainties used to weight each motion measurement in the 3-epoch fit
contain both the individual positional uncertainties as well as the rms
residuals of the transformation. However, the transformation rms is comprised
of the derivation of the mapping solution at each point, and therefore 
includes a contribution from the individual astrometric uncertainties.
Because the astrometric uncertainties and the transformation residuals 
are a function of x,y position on the images, the two components cannot
be separated. The overestimated uncertainties in each motion fit is 
likely caused by this combination of transformation residual error with
the individual astrometric uncertainties in the transformed epochs.
In the standard procedure when deriving the velocity dispersion from 
proper motions, the mean astrometric uncertainty would be subtracted in 
quadrature. As this mean is larger especially in the y-direction than the 
dispersion value, the velocity dispersion cannot be reduced from the 
astrometric uncertainties in this way. The slightly lower median errors
of $xpm_{err,median} = 0.243$ mas/yr and $ypm_{err,median} = 0.229$ mas/yr
would lead to a reduced velocity dispersion of 
$sqrt(\sigma_x^2 - xpm_{err,median}^2) = 0.119$ mas/yr (4.53 km/s) and
$sqrt(\sigma_y^2 - ypm_{err,median}^2) = 0.089$ mas/yr (3.39 km/s), and 
hence a mean 1D velocity dispersion of $4.0 \pm 0.6$ km/s. Nevertheless,
the fact that the mean astrometric uncertainties are larger than the 
measured dispersion values suggests that the 3-epoch sample is just
not accurate enough to provide a realistic dispersion measurement.

%%% removed mass segregation discussion:
%%%
%%% median/mean uncertainties are substantially larger even for stars K < 14 mag!!!
%%% for 25 cluster stars K < 14: stddev(xpm) = 0.165    mean(xpmerr) = 0.210 (!)
%%%                              stddev(ypm) = 0.220    mean(ypmerr) = 0.220 (!)
%%% the values would be 6.3 and 8.4 km/s anyways, and therefore still somewhat 
%%% close to the limits given above.

In summary, an upper limit of the velocity dispersion of $\sim\!\! 10$ km/s is obtained
for the core of the Quintuplet cluster.
The photometric mass in the cluster center was recently measured by Hu{\ss}mann et al.~(2012) 
to be $M\sim 6000\,M_\odot$ for $0.5 < M/M_\odot < 60$ within a radius of 0.5 pc.
Inverting the equation for the virial mass within radius $r$, 
$M_{vir} = 2\cdot r\cdot\sigma_{3D}^2 / G$ within 0.5 pc of the cluster center
(where $G$ is the gravitational constant), the 
expected 3D velocity dispersion would be on the order of $\sigma_{3D} \sim\!\! 5$ km/s, 
and $\sigma_{1D}$ could be as small as $\sim\!\! 3$ km/s. 
Such a low central velocity dispersion would be consistent with the 
measurements in other young, massive clusters such as the Arches ($\sigma_{1D} = 5.7$ km/s
\footnote{This value is deduced from the dispersion measurement of 0.15 mas/yr
for a GC distance of 8.0 kpc.}, Clarkson et al.~2012)
and NGC 3603 ($\sigma_{1D} = 4.5$ km/s, Rochau et al.~2010). Further proper motion epochs 
are therefore required to alleviate the constraints on the derived upper limit.

\subsection{Orbit simulations}
\label{orbitsec}

Orbital simulations were carried out following the prescription in Stolte et al.~(2008).
With the measurement of the 3D orbital velocity and the two spatial coordinates on the 
plane of the sky, the only unknown is the line-of-sight distance to the cluster
(from the Sun), which is represented in the following as the line-of-sight distance between 
the cluster and the Galaxy's center of mass (henceforth called ``line-of-sight distance''
for simplicity). As the cluster is evolved in the gravitational potential of the 
inner Galaxy, no assumption needs to be made about the absolute distance between the 
GC and the Sun. 
As shown in the case of the Arches cluster, orbits at large radii become increasingly
self-similar, as expected at larger distances within the tidal field of the inner Galaxy.
Dramatic changes do occur in the orbital characteristics, and in particular in the closest 
approach of the cluster to the supermassive black hole (Sgr A$^\ast$), if the cluster is
located at small seperations from the center of the potential, 
which implies that its present-day projected 
distance is close to its true distance from the Galaxy's center of mass.
We evolved a set of orbits with line-of-sight distances between -200 pc and 
+200 pc from the Galactic center.
As in the case of the Arches cluster, the Quintuplet was assumed to be a point mass
orbiting in the gravitational potential of the inner central molecular zone. 
The potential consists of the central black hole, the nuclear stellar cluster ($r < 10$pc),
the nuclear stellar disk ($r < 200$ pc), beyond which the flattened potential is 
smoothly transitioned into the potential of the Galactic bar.\footnote{The 
Galactic bar is mesaured to have a pattern speed of $1.9\times \Omega_{circ}$,
where $\Omega_{circ}$ is the local angular rotation velocity of the Milky Way 
(Gardner \& Flynn 2010). For an observer at the solar circle, this pattern speed implies a
rotation period of 124 Myr for the bar, which might influence the orbital motion
for cluster distances beyond 200 pc from the GC. This period is long compared to 
the lifetime and the orbital timescale of each cluster ($< 5$ Myr), such that we
have not incorporated bar rotation in our orbital simulation.}
For a detailed set of parameters and the fit to the enclosed mass, see Stolte et al.~(2008).

\begin{figure*}
%\hspace*{-1cm}
\includegraphics[width=8.4cm,angle=90]{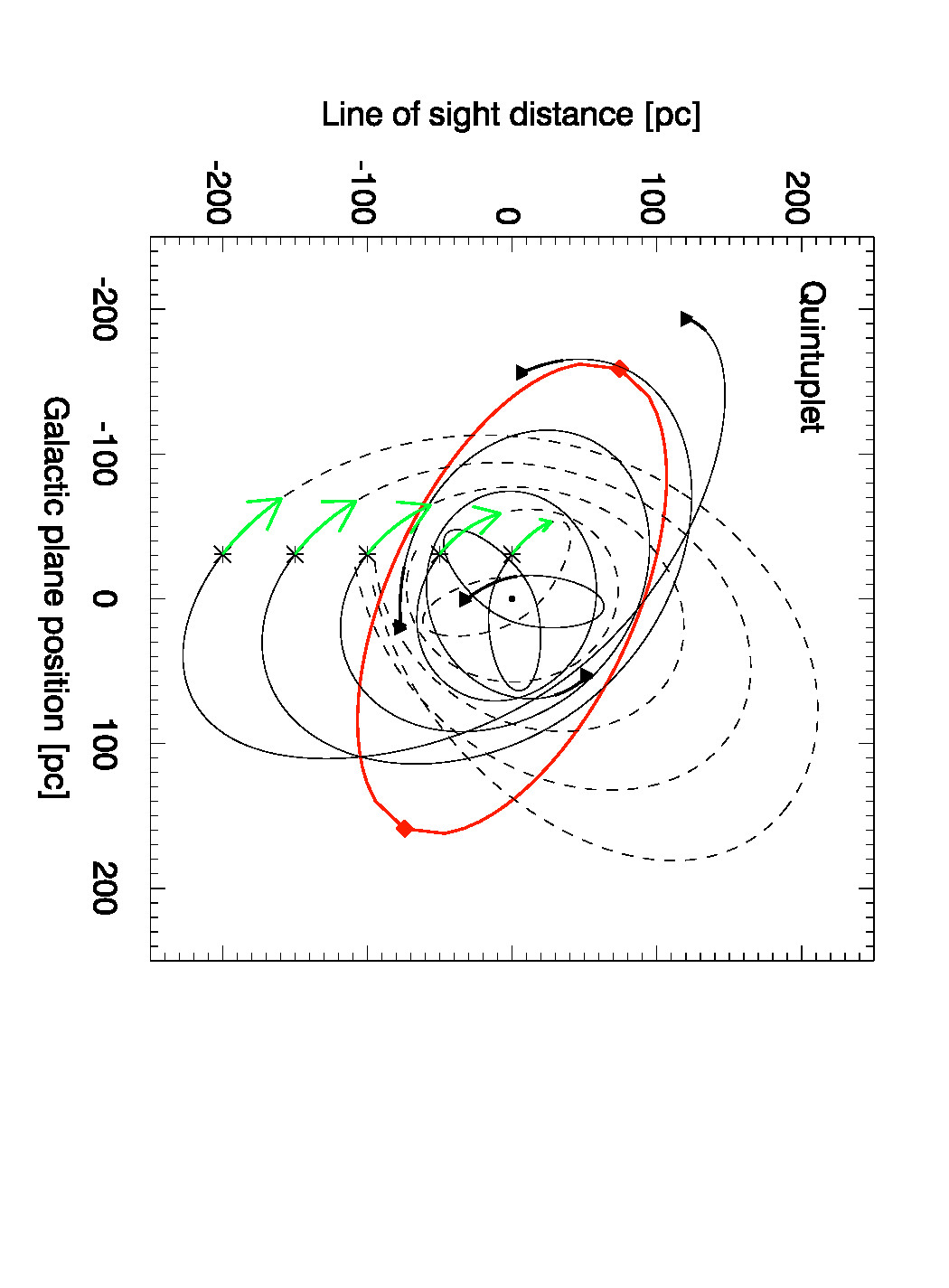}
\hspace*{-3cm}
\includegraphics[width=8.4cm,angle=90]{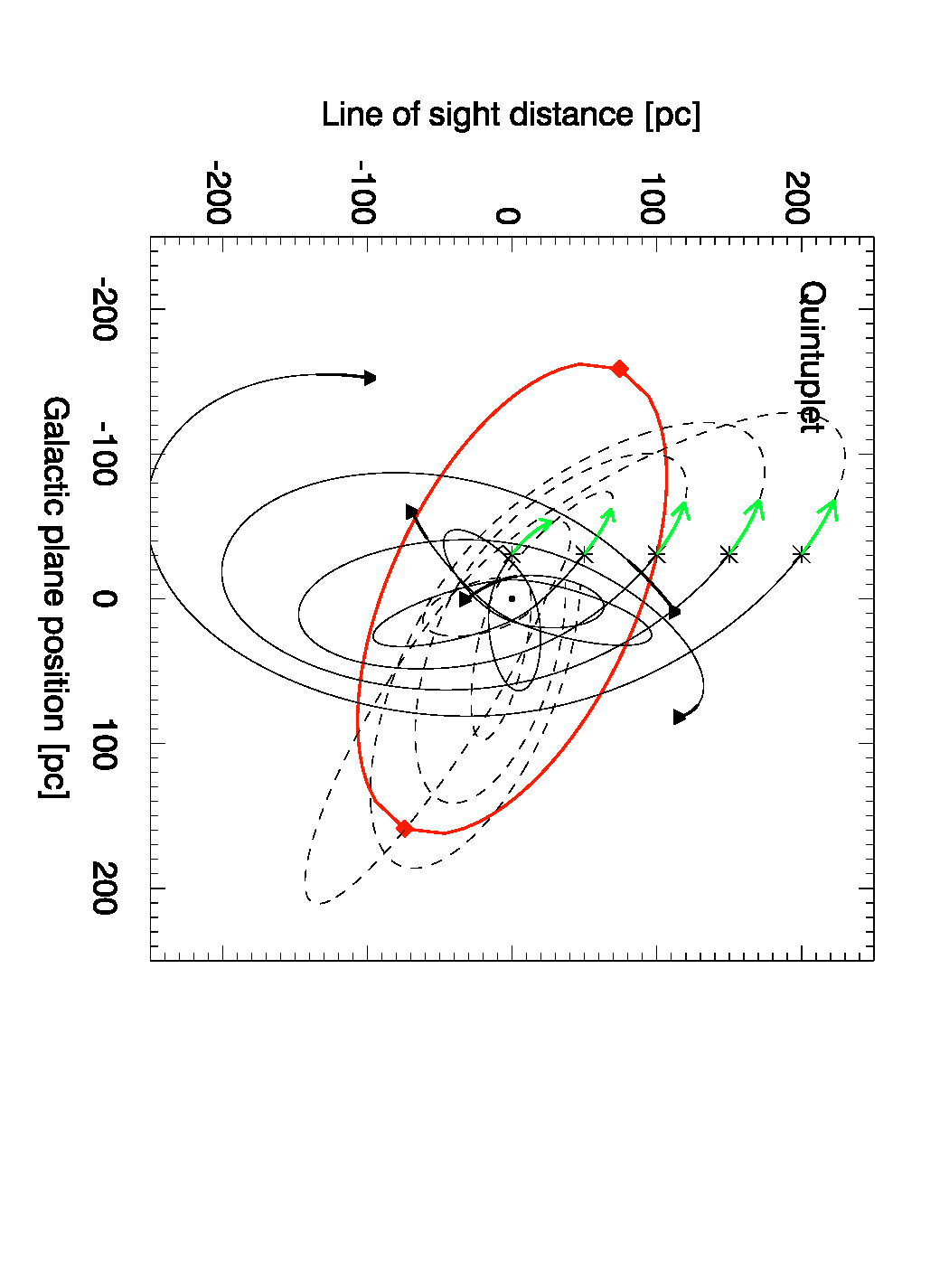}
\caption{\label{orbits}
Orbital simulation of the Quintuplet cluster as a point mass in the 
GC potential. The orbits are partially nested. The supermassive black 
hole (Sgr A$^\ast$) is marked as a black dot at the centre. 
The present-day position of the Quintuplet cluster is shown as an asterisk.
The solid line represents the backward integration to the presumed origin 
of the Quintuplet 4 Myr ago, with each starting point marked as a triangle. 
The dashed line is the projection of the orbit into the future, and the 
current direction of motion is indicated by the green arrow along the orbit. 
Orbits with present-day line-of-sight distances between -100 and -200 pc
({\sl left panel}) suggest a cluster origin close to the Eastern endpoint of the 
x2 orbital zone (shown as an ellipse with the x1-x2 tangent point marked
as a red square). Note that these orbits in front of the GC are 
prograde with the rotation of the bar and of clouds in the inner Galaxy.
For a present-day location of the cluster behind the GC 
(retrograde orbits, {\sl right panel}), no prefered point of origin is observed. 
The location near the endpoint
of the x2 orbital zone is consistent with a formation scenario where gas and 
dust are channeled into the central molecular zone through interaction with 
the bar potential.}
\end{figure*}

The Quintuplet orbital family is calculated using the 3D velocity derived in the 
previous section as the boundary condition for the cluster's present-day motion 
(Fig.~\ref{orbits}). For clarity, the left panel shows the orbits in the case
that the cluster is located in front of the Galactic center today, while orbits for a 
present-day location behind the GC are shown in the right panel. Note that there is weak
evidence from the extinction towards the Quintuplet ($A_{Ks} = 2.35$ mag, Hu{\ss}mann et al.~2012), 
which is lower than the extinction of sources in the vicinity of Sgr A$^\ast$ 
($A_{Ks} = 2.54$ mag, Sch\"odel et al.~2010) and in the Arches cluster ($A_{Ks} = 2.5-2.6$ mag, 
Habibi et al.~2013), that the cluster is situated in front of the GC today.
A location in front of the GC implies that the Quintuplet is on a prograde orbit, which 
is also consistent with the gas velocities in the central molecular zone (e.g., 
Dame et al.~2001, see especially their Fig.~3) and 
with the angular precession of the Galactic bar (Binney et al.~1991). % remark by Jessica

\begin{figure*}
\centering
\includegraphics[width=10cm,angle=0]{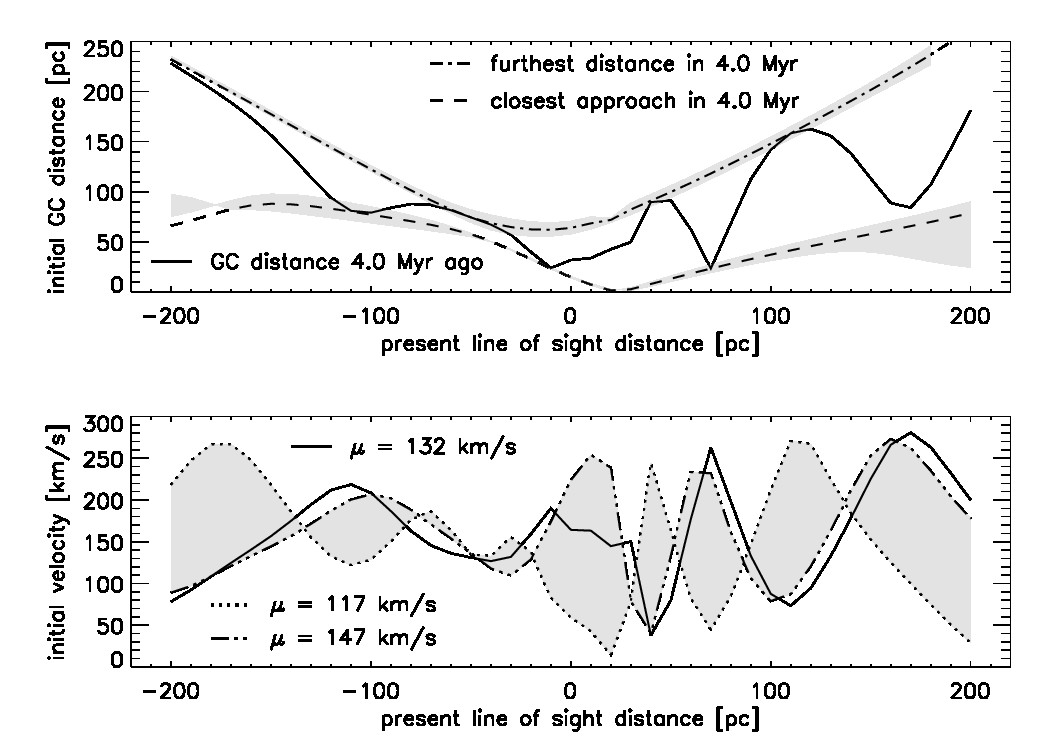} \\
\includegraphics[width=10cm,angle=0]{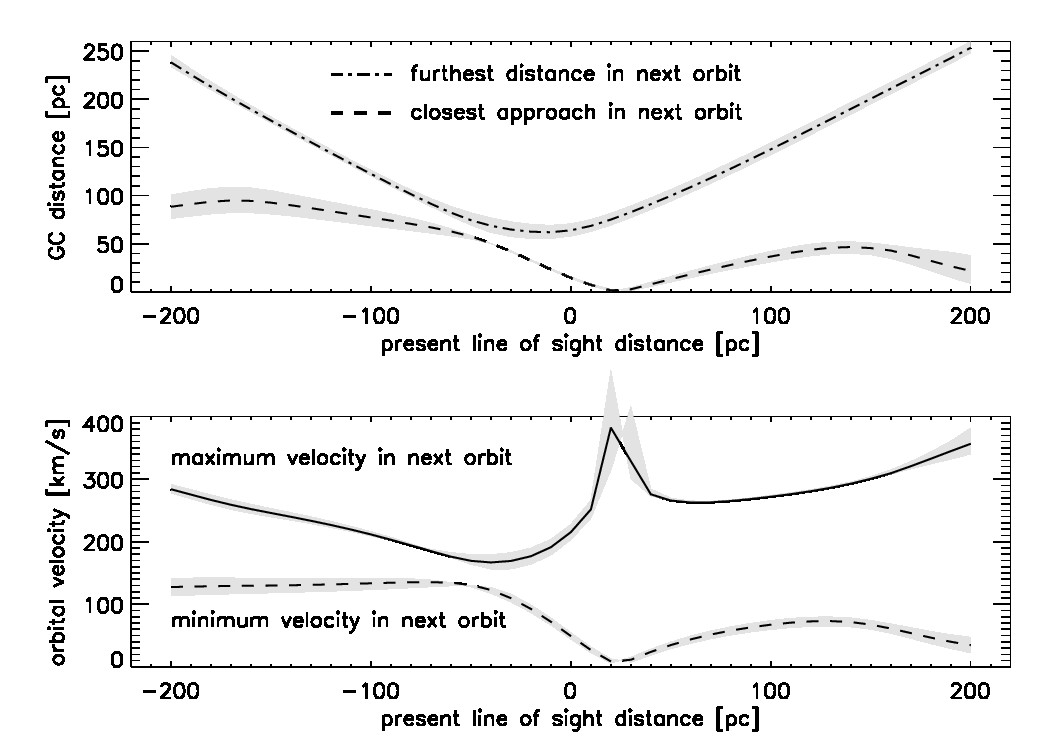} 
\caption{\label{initial} The initial ({\sl top two panels}) and future 
({\sl bottom two panels}) orbital evolution of the Quintuplet cluster. 
The first panel shows the initial distance 
and the furthest and closest approach of the cluster to Sgr A$^\ast$
(i.e.~the center of the gravitational potential),
while the second panel shows the initial orbital velocity 4 Myr ago.
The {\sl bottom panels} display the closest and furthest distances from Sgr A$^\ast$
as well as the minimum and maximum velocities during the next full orbit.
The grey area displays the limits for the 1-sigma proper motion uncertainty of 
$\pm 15$ km/s, and hence for orbits with present-day proper motions of 117 and 147 km/s
(corresponding to 3D orbital velocities of 155 and 180 km/s).
The predictions of the closest and furthest approach
since the cluster's formation suggest that for most line-of-sight distances, the 
Quintuplet has not come closer than its current projected distance of 32 pc into 
the central region during pericenter passage, and will likely not do so in 
the future either. Only if the cluster is located at a line-of-sight distance of 
$20 \pm 20$ pc {\sl behind} the GC today, is it likely to migrate closer than 32 pc 
into the nucleus.}
\end{figure*}

Assuming an age of 4.0 Myr (Figer et al.~1999b, but see also Liermann et al.~2010), 
the orbital motion was integrated backwards in time to the point of the Quintuplet's
expected origin. Within this time, the cluster concluded approximately one orbital
revolution for most present-day line-of-sight distances. Only on the two innermost
orbits would the Quintuplet have completed several revolutions within its present
lifetime. 
The past and future orbits and the cluster's approach to the center of the 
gravitational potential are analysed in Fig.~\ref{initial}.

If the cluster is located within the central molecular zone, $R_{GC} < 200$ pc,
its initial distance from the supermassive black hole ranged from 20 to 230 pc
(Fig.~\ref{initial}, top panel). Only in a narrow range of orbital solutions
was the cluster located close to the supermassive black hole during its first
circumnuclear passage (see dashed line in Fig.~\ref{initial}, top panel).
Especially on the innermost orbits, the natal cloud of the Quintuplet had to 
approach the GC to within less than 100 pc, well inside the central molecular ring.
Even if the progenitor cloud was a member of the central molecular zone, 
it did not follow the Keplerian orbits with moderate velocities as found in 
the central star-forming ring (Molinari et al.~2011).
Instead, the innermost orbits require that the Quintuplet's parental cloud had 
an improbable inward velocity that let to a strong deviation from a circular orbit
(see Fig.~\ref{orbits}).

The uncertainty in 
the proper motion of the cluster, $\pm 15$ km/s, is used to model
the uncertainty in the orbital parameters. Minimum velocity orbits were 
calculated assuming a present-day proper motion of 117 km/s, and maximum velocity
orbits were derived from a present-day proper motion of 147 km/s. 
The present-day 3D minimum and maximum orbital velocities then correspond
to 155 and 180 km/s when combined with the radial velocity of 102 km/s.
The uncertainty
in the closest and furthest approach to Sgr A$^\ast$ and the initial cluster 
velocity at its presumed birth time are displayed as grey areas around the 
lines, which represent the orbital parameters for the measured proper motion 
of $\mu = 132$ km/s. 
It is particularly striking that the closest and furthest approach from Sgr A$^\ast$ 
do not change substantially given the uncertainty in the velocity measurement.
This is true for both the initial orbit (integrated backwards to the cluster's 
presumed origin) and the next revolution around the GC (see Fig.~\ref{initial}).
The maximum velocity in the next (future) orbit is most sensitive to the 
velocity uncertainty on the few innermost orbits where the cluster is proceeding 
very close to Sgr A$^\ast$, as expected. The largest uncertainty is observed in 
the initial velocity. The initial velocity depends severely on the exact point
of the cluster's origin. If the velocity is read off slightly earlier along 
the orbit (for a faster orbit or a slightly older cluster age, see also Fig.~\ref{orbits}), 
the cluster will have moved to a different location in the gravitational potential.
Likewise, if the initial velocity is read off closer to its present position
(for a slower orbital motion or a slightly younger cluster age), the velocity
can have changed substantially. As a consequence, the grey areas in the second
panel of Fig.~\ref{initial} partially represent a phase shift. For the minimum
velocity orbit with $\mu = 117$ km/s the slow cluster motion implies a stronger 
influence from the gravitational potential, and hence more variation in the 
velocity at each position. In summary, while the exact value of the initial
velocity of the Quintuplet is sensitive to the uncertainty in the 
present-day proper motion, both the minimum and maximum velocities as well as
the closest and furthest approach from the center of the Galaxy are robust 
against the measured proper motion uncertainty.

\subsection{The Quintuplet's approach to SgrA$^\ast$}

Early after the discovery of the central clusters, Gerhard (2001) suggested that
inspiraling clusters might fuel the young stellar population surrounding the 
supermassive black hole. Follow-up simulations by Kim \& Morris (2003) and 
Portegies Zwart et al.~(2002) suggested that clusters need be on eccentric orbits
or need extreme properties in terms of cluster density and mass in order to 
deposit stars near Sgr A$^\ast$. The eccentric orbits suggested by our simulations
for the Quintuplet might provide the necessary setup for the cluster to closely
approach the nucleus.
In order to evaluate how close the Quintuplet could possibly have come to the central parsec, 
the properties of the next full orbit after the present-day location are also shown in 
Fig.~\ref{initial}. There is only one extreme case where the cluster would migrate
into the inner few parsecs. For a line-of-sight distance of 20 pc {\sl behind} the GC today, 
the cluster could have come as close as 2 pc to Sgr A$^\ast$ during its 4 Myr 
lifetime (at an age of 1.1 Myr), and it would 
again pass Sgr A$^\ast$ with a minimum distance of $\sim 2$ pc with a 3D orbital velocity of 
380 km/s at an age of 7.2 Myr on this orbit. 
According to these simplistic simulations, the cluster would need
to be at a line-of-sight distance between 10 and 40 pc behind the GC today in order 
to reach the central 10 pc around Sgr A$^\ast$. The high velocity of 300-400 km/s of the cluster 
during the passage through the inner few parsecs limits the number of stars that could have 
been tidally stripped. Even in the case of the most eccentric orbit, the nuclear population 
of more than 80 young, early-type stars residing in the central parsec today 
(Bartko et al.~2010, Do et al.~2013)
could not easily be explained by tidal stripping from the Quintuplet cluster.
Given its likely location in front of the GC today, an interaction between the 
cluster and the nuclear population or the supermassive black hole seems quite unlikely.

\subsection{Orbital inclination}

\begin{figure*}
\centering
\includegraphics[width=14cm]{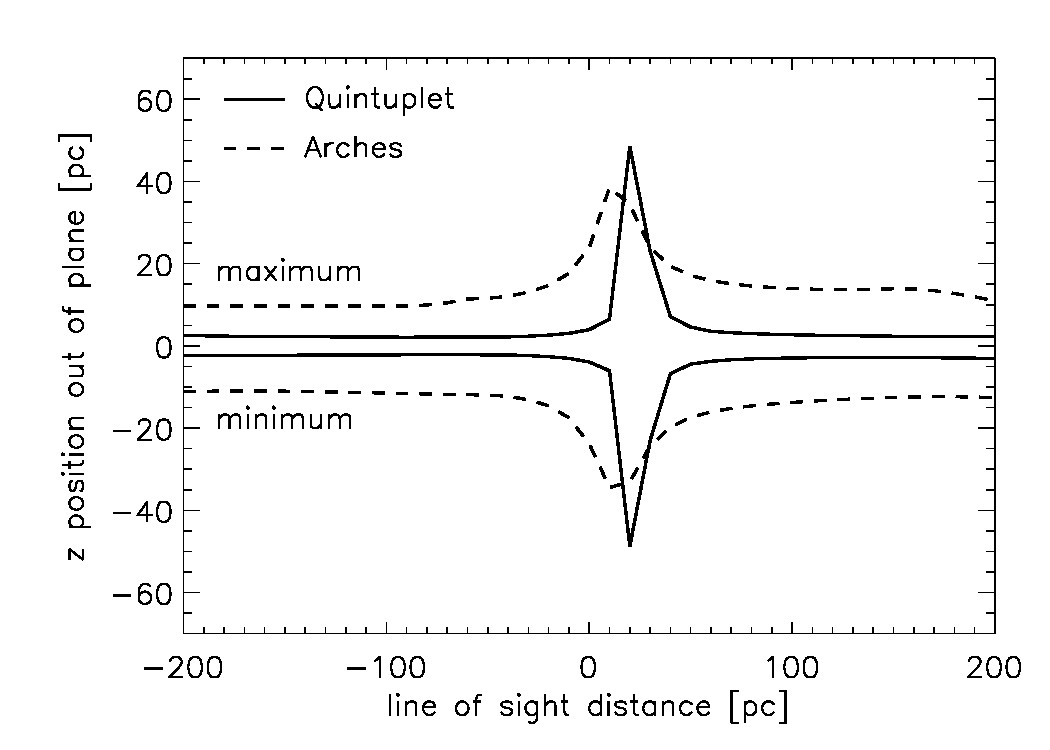}
\caption{\label{zminmax}
The minimum and maximum elevation below and above the Galactic plane
reached by the Arches and Quintuplet clusters, as a function of their
respective GC distance along the line of sight. The minimum and maximum
were calculated backwards in time for the age of each cluster, 2.5 Myr 
and 4.0 Myr for Arches and Quintuplet, respectively, and forwards in time
until 8 Myr from their present location. The large out-of-the-plane motion
indicated for orbits at line-of-sight distances of 0 to 30 pc behind the 
GC today are caused by the close approach of both clusters to the Galactic 
nucleus on these orbits.}
\end{figure*}

In Fig.~\ref{zminmax}, the minimum and maximum elevations above the 
Galactic plane are compared for both the Arches and the Quintuplet clusters. 
These extrema are derived
for the entire duration of each cluster's orbit from its origin (2.5 and 4.0 
Myr ago, respectively) until 8 Myr into the future from their present-day 
location.
For most orbits, the Quintuplet cluster remains closer to the Galactic plane
than the Arches. This results directly from the current location of the Quintuplet
very close to the disk plane, and the negligible out-of-the-plane motion component. 
The typical out-of-the-plane ($z$) motion is similar for all orbits where the 
Quintuplet remains far from the nucleus. Here, the cluster stays within $\pm 5$ pc
of the Galactic plane for most orbits, and in some cases at line-of-sight distances 
of 100-200 pc behind the GC today reaches a maximum $z$-elevation of 12 pc. 
Only for orbits where the cluster enters the zone of influence of the nuclear 
cluster and black hole, migrating to distances of a few parsecs from the gravitational
center, is the orbit heavily perturbed. In these cases, the orbital motion can 
reach altitudes of as much as 60 pc below and 40 pc above the Galactic plane
(Fig.~\ref{zminmax}). Exclusively at present-day line-of-sight distances of 
20 to 30 pc behind the GC, does the Quintuplet penetrate as close as 2 to 4 pc 
into the gravitational center. The high velocities cause the cluster to 
experience sling-shot perturbations out of the Galactic plane.

This general pattern is similar for the Arches orbit (Fig.~\ref{zminmax}).
Yet, the Arches's current position 10 pc above the plane causes the 
maximum elevation to remain at 10 to 20 pc for most orbits, such that the 
cluster crosses the Galactic plane multiple times during one orbit.
Even with the revised, and slightly lower, 3D orbital velocity of 172 km/s,
which facilitates the inward motion of the cluster,
the Arches never penetrates the inner 5 pc of the nucleus.
The Arches therefore does not experience equally dramatic sling-shot
perturbations, and stays within distances of $\pm 40$ pc above and below
the Galactic plane during all orbits up to the considered timescale of 8 Myr.

\section{Discussion}
\label{discsec}

\subsection{Deviation from a circular orbit}

\begin{figure*}
\hspace*{1.4cm}
\includegraphics[width=14cm]{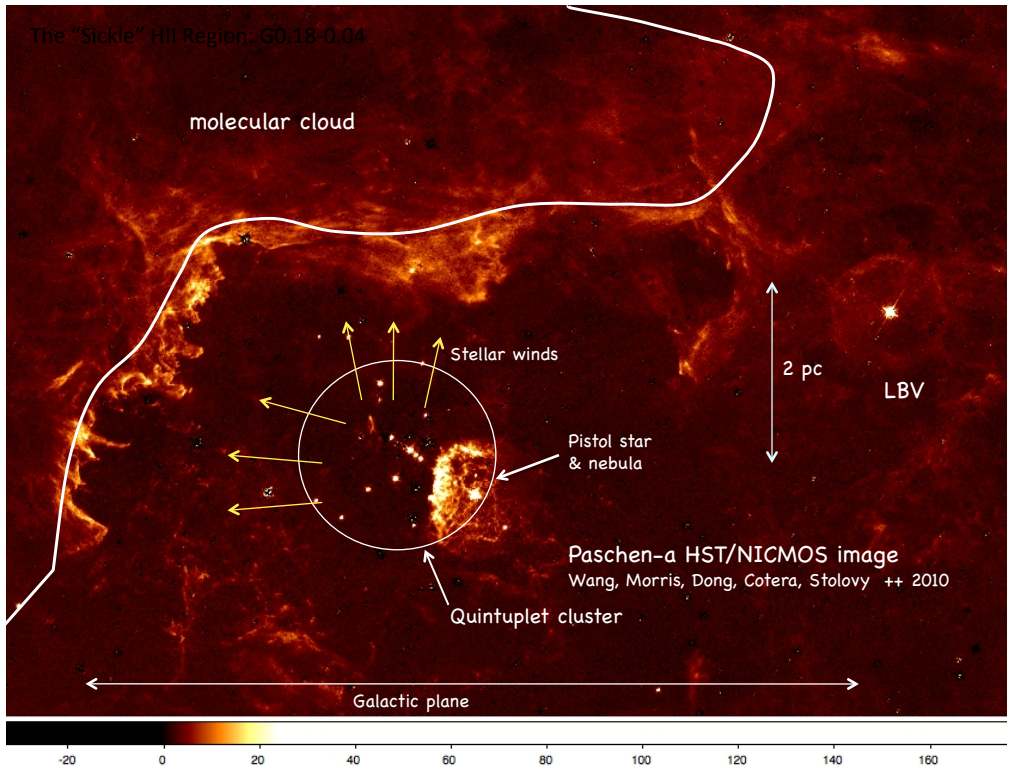}
\caption{\label{sickle} 
Paschen $\alpha$ image of the Quintuplet and Sickle cloud region
from the HST/NICMOS Paschen $\alpha$ survey (Wang et al.~2010, 
Dong et al.~2011). 
Note that because stellar continua have been subtracted from this image, 
as described by Dong et al.~(2011), the Quintuplet stars evident in this 
image are only those stars that show Paschen-alpha emission lines.
The direction of motion of the Quintuplet cluster
is parallel to the Galactic plane indicated by the arrow at the bottom,
such that the cluster moves into the Sickle cloud to the left.
The ionisation rim suggests that the high-mass population of the 
Quintuplet evaporates and ionises the cloud through winds and the
intense UV radiation field.}
\end{figure*}

The 3D orbital velocity of the Quintuplet cluster appears high in comparison to the 
circular velocity at its projected distance of 31 pc from the GC. Assuming the 
enclosed mass estimates from Launhardt et al.~(2002), the circular velocity at
$R_{GC} = 31$ pc is only 90 km/s for an enclosed mass of $M_{enc} = 6\times 10^7\,M_\odot$,
and stays below 150 km/s out to a galactocentric radius of 100 pc
($M_{enc} < 7\times 10^8\,M_\odot$). 
Between 100 and 200 pc distance from the GC, the Keplerian velocity
$v_{circ}$ theoretically increases
to 190 km/s according to an increase in the enclosed mass from $M_{enc} = 6\times 10^8\,M_\odot$ 
at 100 pc to $M_{enc} = 2\times 10^9\,M_\odot$ at 200 pc. Note, however, that such
velocity values are not measured in objects located in the central molecular zone.
The study of OH/IR stars by Lindqvist et al.~(1992) constrained terminal radial 
velocities to less than 120 km/s at all radii (see also the detailed discussion 
in Stolte et al.~2008). Likewise, radio surveys of dense
clouds in the inner central molecular zone indicate velocities below 120 km/s
for the x2 orbital family (Dame et al.~2001, their Fig.~3, see also Binney et al.~1991).
From this observational evidence, a 3D orbital velocity of 167 km/s is substantially 
higher than maximum radial velocities observed in both clouds and stars in the 
central molecular zone. The simulations of the cluster orbit support the expectation 
that the motion of the cluster is not consistent with a circular orbit.
One could argue that the cluster might be located several hundred parsecs in front
of the Galactic center. However, the recently obtained HST/NICMOS Paschen alpha 
survey of the Galactic center (Wang et al.~2010, Dong et al.~2011) 
clearly shows strong interaction between the massive cluster stars and 
the nearest cloud (Fig.~\ref{sickle}). The pillars and fringes
at the cloud edge in front of the cluster's motion indicate that the ionising
radiation and wind pressure from the Wolf-Rayet cluster members is eating into 
the cloud towards which the cluster is presently moving. 

Three-dimensional hydrodynamic simulations of clumpy cloud surfaces exposed 
to ionising radiation from a single nearby O-type star result in pillared structures 
on timescales of 2 to $4\times 10^5$ yrs (Mackey \& Lim 2010). Comparing the 
radial velocity of the Quintuplet, $v_{rad} = 102$ km/s, with the cloud velocity
of $20-40$ km/s at the location of the Sickle (Molinari et al.~2011, see their Fig.~4),
the cluster would move 15 pc for a relative velocity of $\Delta v_{rad} \sim 80$ km/s
within $2\times 10^5$ yrs. However, the Quintuplet is located only 2-3 pc from 
the illuminated cloud rim in projection at the present epoch.
The higher radiation and wind pressure of $\sim 100$ O-type 
and Wolf-Rayet stars (Liermann et al.~2009, Hu{\ss}mann et al.~2012) 
might accelerate the carving of the pillars, 
and the fact that no ionised rims are observed behind
the direction of motion of the Quintuplet, where the cluster might have cleared
its path, corroborates the suggestion that
the cluster is moving towards the ionisation rim. The increasing flux incident
on the cloud rim during the cluster's approach might cause additional instabilities
aiding in the formation of the pillars. Such a scenario was suggested for 
the individual, rapidly moving star $\xi$ Per (Elmegreen \& Elmegreen 1978), 
and will be enhanced for the large number of high-mass stars in the Quintuplet.
These authors already suggested a shortened formation timescale for pillared 
structures in the case of relative motion between the ionising star and the 
cloud, accelerated further by instabilities forming at the cloud rim.
It would be interesting 
to probe whether the cluster-cloud interaction is capable of triggering 
the next generation of star formation. Right now, however, there is no 
direct evidence for young stellar objects in the compressed cloud.
The interaction provides stringent evidence that the Quintuplet cluster is 
indeed moving through the CMZ, implying a location at a radial distance of 
less than 200 pc from the GC at the present time and a non-circular motion
of the Quintuplet with significant differences between the cluster's 
apocenter and pericenter passages.

\subsection{Comparison with the Arches cluster}

\begin{figure*}
%\hspace*{-1cm}
\includegraphics[width=8.4cm,angle=90]{quin_132_before_orbits.jpg}
\hspace*{-3cm}
\includegraphics[width=8.4cm,angle=90]{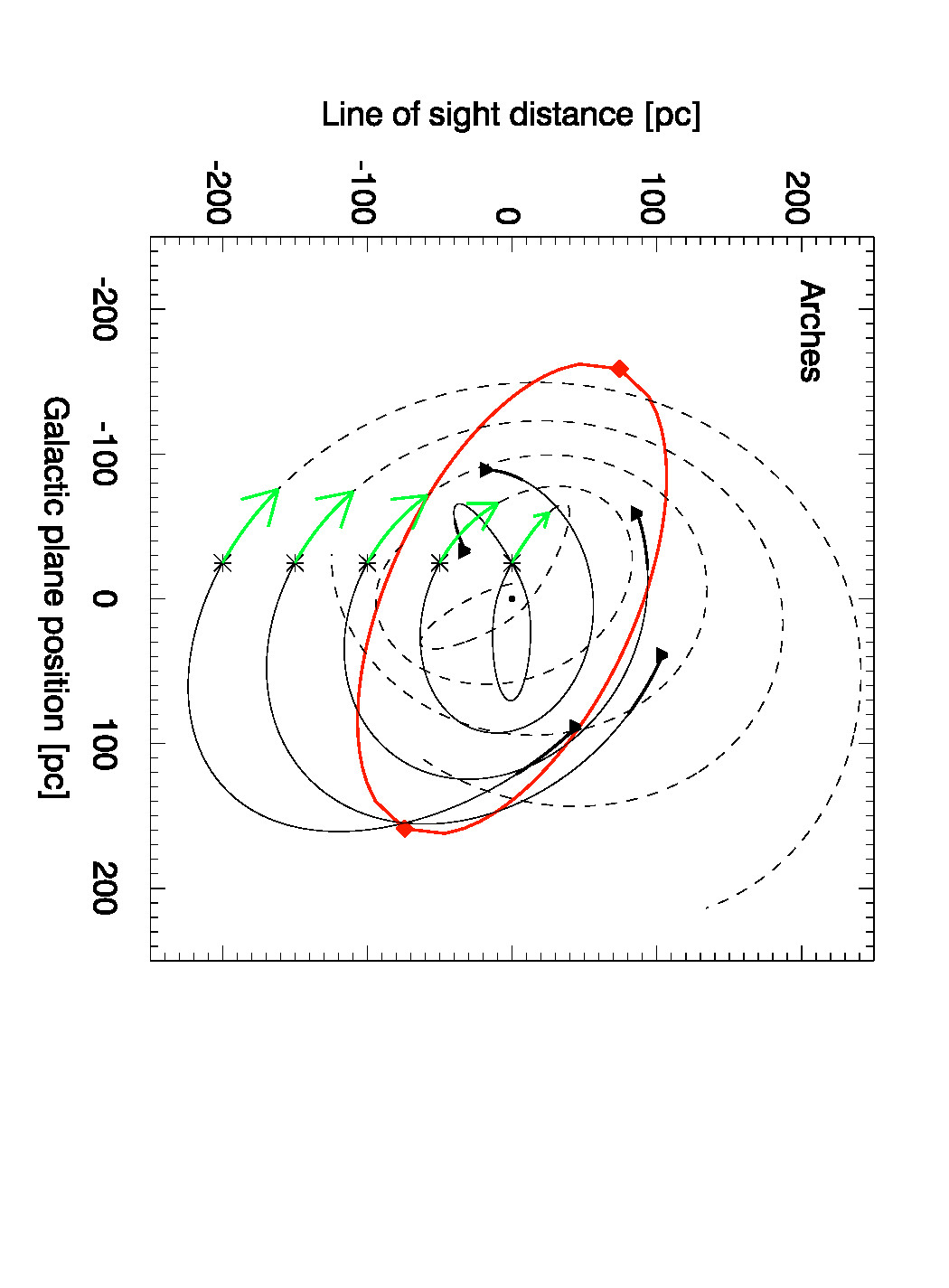}
\caption{\label{archquin_orbits}
Comparison of Quintuplet and Arches orbital simulations. The {\sl left panel} 
is identical to the left panel in Fig.~\ref{orbits}, while the {\sl right panel} 
displays the modelling of the Arches orbit adapted from Stolte et al.~(2008)
for the revised orbital velocity of 172 km/s (Clarkson et al.~2012). 
Labels are as in Fig.~\ref{orbits}, and the view is again from above the Galactic plane. 
The asterisks display the present-day positions of both clusters on each 
respective orbit.
For several line-of-sight distances, both clusters have their possible origin near 
the outermost x2 orbit indicated by the solid ellipse.
The possible origins of the Quintuplet cluster 
cover a wider spatial range than the origins of the Arches. 
It is intriguing that a similar origin can be found for the Arches 
at an age of 2.5 Myr and the Quintuplet at an age of 4 Myr at all.
The longer timescale for the Quintuplet to reach a similar position in space
is caused by the more elliptical orbits, causing the Quintuplet to reach their 
common present-day location at a later age.}
\end{figure*}

A revised version of the Arches orbits for a proper motion of 172 km/s for 
the cluster with respect to the field (Clarkson et al.~2012) is shown in 
Fig.~\ref{archquin_orbits}. 
Only orbits where the Arches is located in front of the GC today are directly 
compared with the respective orbits of the Quintuplet cluster.
The orbits where the Arches cluster is behind the GC today are very 
similar albeit less chaotic and even more regularly nested than the orbits of
the Quintuplet cluster shown for a location behind the GC today in Fig.~\ref{orbits}.
For a location behind the GC today, both clusters would have to be on retrograde
orbits, and orbiting against the rotation direction of the Galactic bar.
However, the ionisation rim near the Quintuplet along the Galactic plane
in the direction of motion 
as well as the apparent interaction of the Arches moving into a nearby cloud
observed in the radio regime (Lang et al.~2003) suggest that both clusters 
are comoving. A formation scenario for massive clusters on retrograde orbits 
would be even harder to find.
While the Arches velocities suggest a family of self-similar, nested orbits,
the Quintuplet orbital family displays more intersection points.
It is striking that especially the approach of the Arches and Quintuplet clusters
to the supermassive black hole is very sensitive to the differences in their 
3D orbital velocities.
In the case of the Arches, the simulations forced us to conclude that the cluster is not 
able to reach the central parsec
before it disperses into the GC field ($r_{GC} > 5$ pc at all times and for all line-of-sight
distances, see also the discussion in Stolte et al.~2008).
The particular motion vectors of the Quintuplet, on the other hand,
and its location on the Galactic plane, promote extreme orbits for line-of-sight
distances of 20-30 pc behind the GC today, where the cluster approaches the center of 
gravity to within 2-4 pc at closest approach. Yet, as discussed above, this 
would require the Quintuplet to be on a retrograde orbit.

The fact that both the Quintuplet and the Arches cluster appear far from a 
circular orbit solution has severe implications for their dynamical evolution. 
N-body simulations treating self-consistently the stellar component of the 
inner Galaxy and the internal dynamics of a dense star cluster have 
recently shown that clusters on eccentric orbits develop chaotic tidal tails (Fujii et al.~2008,
see especially their Fig.~2).
According to  Kim \& Morris (2003) and Fujii et al.~(2008), 
clusters on eccentric orbits migrate faster towards the 
center than the clusters in earlier circular orbit models (e.g., Kim et al.~2000, Portegies Zwart 
et al.~2002). The development of chaotic tidal 
tails additionally implies that tidally stripped stars can end up in positions far away from 
the position of the cluster orbit, as might be evidenced in the population of dispersed
Wolf-Rayet stars in the GC (Mauerhan et al.~2010b, M. Habibi et al.~2014). 

Because of the more chaotic behaviour of the Quintuplet orbital family as compared
to the Arches orbits and because of its older age, the location of the Quintuplet's 
origin is less constrained. For the Arches, the orbit simulations suggest that 
the cluster likely emerged toward the upper left quadrant of the motion plane 
or close to the x2 orbital family (red ellipse in Fig.~\ref{archquin_orbits}). 
A similar origin close to the outermost x2 orbit is likely for the Quintuplet if it is 
located at a line-of-sight distance of -50 to -200 pc in front of the GC today. 
Especially near a present line-of-sight distance of -150 to -200 pc, 
the origin of the Quintuplet was close to the tangent point 
of the x2 orbital zone indicated as the solid ellipse in Figs.~\ref{orbits} and
\ref{archquin_orbits}.
As we already noted in the discussion of the Arches orbit (Stolte et al.~2008), 
a location near the outermost x2 orbit is consistent with a formation region between 
the x1 and x2 orbits of the Galactic bar. This transition region is located at the outer edge 
of the central molecular zone, where instreaming and circumnuclear clouds are prone 
to collisions. This formation scenario is also consistent with both clusters
being on prograde orbits, which implies a location of both clusters in front of 
the GC today.

%%%
Both on orbits located further inward and on orbits where
the Quintuplet is located behind the GC today, the cluster would have emerged from 
its natal cloud at vastly different locations from the most likely place of birth
of the Arches.
One should keep in mind, however, that these conclusions are sensitive to the assumed
Quintuplet cluster age of 4.0 Myr 
(one to a few orbital periods for outer and inner orbits, respectively), 
and that the embedded phase during which the cloud fragmented
and the cluster contracted into its current dynamical state are not accounted for.
Most of the orbits where the cluster is located in front of the GC today, with the 
exception of the innermost cases, would be consistent with a formation locus near
one of the x1/x2 transition points if the cluster were slightly older ($\sim 0.5$ Myr).
For the case that both clusters have emerged at one of the end-points of the 
x2 orbital family (see Fig.~\ref{archquin_orbits}), a consistent formation scenario 
of these two massive clusters is presented in the next section.

\subsection{The origin of the GC clusters}
\label{clusorigin}

The close proximity of the Arches and Quintuplet clusters in the central molecular zone,
at a projected spatial separation of only 12 pc from each other, and their similar motion 
parallel to the Galactic plane, raise the question of whether these two clusters have a
common origin, or, at least, have originated in a similar fashion.
It is striking that the Arches and the Quintuplet are the only 
{\sl compact, massive} young clusters in the central region outside the nucleus.
Previously, with cluster ages of $2.5 \pm 0.5$ Myr for the Arches and $4 \pm 1$ Myr for 
Quintuplet, the age discrepancy was considered too large for both clusters to 
originate from the same molecular cloud. At the time when the Arches must have formed,
the Quintuplet's most massive stars would already have excavated the native
cloud substantially, such that the formation of a second, equally massive twin would seem
unlikely. This argument is somewhat diminished by the recent age dating of five hydrogen-rich
Quintuplet WN stars by Liermann et al.~(2010), suggesting that these stars have ages between 
2.4 and 3.6 Myr, and that the main sequence OB population is in the age range of 3.3-3.6 Myr
(Liermann et al.~2012). Such ages would bring both clusters much closer in their evolution, and 
hence would render a common origin more likely. Nevertheless, one has to bear in mind that
Wolf-Rayet evolutionary models still harbour significant uncertainties.
In addition, the luminosity of these objects might be biased if the WNs are located 
in binary systems (Liermann et al.~2010), as is the case for the Quintuplet WC members 
(Tuthill et al.~2006). In this case, the WNs would be affected by binary mass transfer 
and hence appear rejuvenated as compared to the main sequence population (see Liermann et 
al.~2012 for a discussion).
Despite the known uncertainties in the age determination of Wolf-Rayet stars, the fact
that the carbon-rich variety of WC stars is already heavily present in the Quintuplet cluster, 
and in fact led to its early discovery by Okuda et al.~(1990) and Nagata et al.~(1990), 
suggests a more evolved evolutionary stage than the Arches population.
The WC evolutionary stage is expected to follow the WN phase after a brief 
hydrogen-free WN period (Crowther 2007, Martins et al.~2008). Especially the dust-rich 
interacting-wind binaries resolved into spectacular spiral patterns by Tuthill et al.~(2006)
are entirely absent in the Arches cluster. Indeed, spectral analysis suggests that 
all of the Arches WN population are of types WNh6-7 close to or on the high-mass 
main sequence, with strong hydrogen emission line spectra (Martins et al.~2008). 
Hence, while the exact age difference between the Arches 
and Quintuplet clusters is still not well established, there is strong evidence that 
both clusters did not form at the same time, but at least 0.5 to 1 Myr apart.

Even if the Arches and Quintuplet did not emerge from the same cloud, we may be able to 
identify a consistent formation scenario for both clusters that might explain their
similar orbits and their present-day proximity. In extragalactic circumnuclear rings, 
the endpoints of the inner bar were shown recently to bring forth star clusters at 
regular intervals. Especially striking 
is the case of the spiral galaxy NGC 613, where a string of circumnuclear clusters with 
ages between 2-10 Myr is found emerging from both fueling points of the inner 
ring of circumnuclear clouds (B\"oker et al.~2008). This ring of young star clusters displays 
a continuous age sequence of clusters being formed every 2-3 Myr on each side of the inner bar.
At these points, N-body simulations show strong cloud-cloud collisions after gas has 
efficiently streamed in through the outer bar potential (Regan \& Teuben 2003, 
Rodriguez-Fernandez \& Combes 2008, Kim et al.~2011, see also the early gas flow models 
by Athanassoula 1992). In these simulations, dust lanes form along the inner x1 orbit,
and meet infalling clouds in the circumnuclear ring at the contact points between the 
x1 and x2 orbital families. The efficiency of this mechanism suggests that this scenario 
should be present in most, if not all, gas-rich barred spirals, at least in the case that
an inner bar, or circumnuclear ring, is formed from the in-streaming material.
In fact, numerous galaxies are found to have circumnuclear rings lined with young 
star clusters (Mazzuca et al.~2008, and references therein),
with 50\% of their sample displaying a systematic age sequence. As pointed out in the 
detailed discussion of Mazzuca et al.~(2008), the youngest H{\sc ii} regions frequently
emerge near the contact points between the observed dust lanes
and the circumnuclear rings, as expected from the simulations.

In the Milky Way, the existence of an inner bar is still disputed, as such characteristics
are much more difficult to corroborate by observations through the dense Galactic plane
(see the discussion in Rodriguez-Fernandez \& Combes 2008). Nevertheless, the circumnuclear
ring of molecular clouds observed in extragalactic systems can be identified with the 
central molecular zone, which is comparable in its spatial dimensions ($r \sim 200$ pc) to 
the circumnuclear rings in other spiral galaxies (see Table 1 in Mazzuca et al.~2008). 
In addition, the presence of a star-forming ring was recently postulated by 
Molinari et al.~(2011), and several of the dense, compact clouds along this ring 
are expected to form massive clusters comparable to the Arches and Quintuplet
systems (Longmore et al.~2012, 2013). 
If the large outer bar is responsible for fueling the central molecular zone,
the points where infalling clouds are colliding with the existing circumnuclear ring at
a radius of about 200 pc would be a pre-destined place for cluster formation. 
In this scenario, the most massive clusters would form in pseudo-regular intervals, whenever
a cloud having a sufficient mass reservoir collides with material in the central molecular zone.
This would naturally explain how both clusters could have inherited similar velocities
and orbital motions. 
In addition, if an inner bar exists, the endpoints of this bar would 
be the places where gas is channeled in. As seen in extragalactic systems, 
in-streaming clouds collide with the circumnuclear ring of molecular clouds at these 
contact points (Mazzuca et al.~2008), rendering them preferred 
cluster-formation loci. Following the discussion in Molinari et al.~(2011)
and Longmore et al.~(2013), Sgr B2 and Sgr C represent the overdensities 
at the endpoints of the {\sl star-forming ring}, which would then trace 
the cluster formation loci in the Milky Way's CMZ.
The approximate starting points of the orbits (thick triangles in Fig.~\ref{orbits} and 
Fig.~\ref{archquin_orbits}) suggest that both clusters formed at different loci near the 
outer boundary of the CMZ.
In this case, the spatial proximity of the clusters would be a coincidence, 
but would be reconciled by the fact that they formed at similar Galactocentric radii, 
and hence again are found on similar orbits. 

While the recent observations support cluster formation from dense CMZ clouds, 
it remains an open question how these clouds migrate into the central region.
One possibility to channel gas into the CMZ, and possibly to smaller radii,
is provided by models of nested bars. A nested bar model with a 3.5 kpc outer
bar and a nested 150 pc inner bar can explain several of the kinematic
features in the longitude-velocity diagram towards the inner Galaxy 
in a self-consistent gas flow model (Rodriguez-Fernandez \& Combes 2008,
see also Namekata et al.~2009). 
In particular, the $l=1.3\degr$ molecular cloud complex corresponds in this 
model to the far branch of an inner two-arm spiral pattern that channels gas
from the 300-800 pc HI ring into the central molecular zone. 
The lack of a counterpart of the 
$l=1.3\degr$ complex is interpreted as only one of the two theoretically 
predicted spiral arms being active at the present time. Intriguingly, this 
complex is located in the same quadrant where the Quintuplet and Arches orbit
simulations suggest a possible formation locus of these star clusters.
The projected distance of the complex from the Galactic center is 180 pc,
consistent with the formation of the star clusters at the edge of the x2 
orbital zone. 

The present set of large-scale simulations of the fueling of the central molecular zone 
in the framework of barred potentials cannot trace particles down to the spatial scales
required to probe star formation. Due to efficiency and the large spatial scales
involved, self-gravity and feedback effects have so far been ignored. 
Recent high-resolution simulations of the inner few hundred parsecs conducted by 
Kim et al.~(2011) include self-gravity and feedback for the first time. 
It will be intriguing to see whether these N-body simulations of infalling cloud
particles can account for the formation of massive clusters on non-circular orbits, 
such as suggested from the Arches and Quintuplet orbital solutions.

\section{Summary and Conclusions}
\label{sumsec}

We obtained the proper motion of stars in the Quintuplet cluster and the 
field population along the line of sight towards the Galactic center from three 
epochs of high-spatial resolution near-infrared imaging with Keck and VLT.
The relative proper motion of the cluster with respect to the field is derived to be
$132 \pm 15$ km/s from two-dimensional fits to the proper motion density distribution. 
The motion axis is consistent with the Quintuplet cluster moving along the Galactic 
plane, while the motion perpendicular to the plane is negligible.
Combining the proper motion with the known radial velocity leads to a 3D orbital 
motion of $167 \pm 15$ km/s, with the cluster moving outwards from the Galactic center
and receding from the Sun. This orbital motion is surprisingly similar to the 
orbital motion of the Arches cluster (Stolte et al.~2008, Clarkson et al.~2012).

Simulations of the cluster orbit in a multi-component potential of the inner 
Galaxy suggest that the orbit of the cluster is non-circular if the cluster 
is located in the central molecular zone ($R_{GC} < 200$ pc). 
A revised orbit solution for the Arches cluster is also presented.
Tracing the orbital motion of both clusters back for 2.5 and 4.0 Myr to the 
possible points of origin,
a common cluster formation scenario for the Arches and Quintuplet clusters is 
identified if the Quintuplet is at a line-of-sight position in front of the GC today.
In this scenario, the cluster-forming clouds are located inside the inner 200 pc 
of the CMZ and at the far side of the Galactic center, 
towards increasing Galactic longitudes. When compared to simulations fueling
the central molecular zone with nested bar potentials, this location is intriguingly 
similar to the tentative collision point of the rearward spiral arm of the inner bar. 
Today, the $l=1.3$\degr\ molecular complex has accumulated a total mass of 
$\ge 2\times 10^5\,M_\odot$ at this position (Oka et al.~1998). 
Although there are presently no 
signposts of star formation in the form of HII regions in this complex, it is 
conceivable that the next compact, massive cluster will be forged in the near 
future at this exceptional locus.
Further simulations might provide insight as to whether the collisional properties
of the infalling gas into the central molecular zone are sufficient to explain the 
high velocities of the Arches and Quintuplet clusters, and their non-circular orbits. 

The current velocity measurement provides the basis for a new set of simulations
to reconstruct the tidal losses over each cluster's lifetime. Especially the diffusion
of high-mass stars from the young population might have given rise to the population
of apparently isolated young, high-mass stars in the GC. These simulations would 
solve the mystery of apparently isolated high-mass star formation in the GC
environment.
Observationally, increasing the proper motion accuracy with more epochs over a longer
time baseline
will lead to the measurement of the internal velocity dispersion of the Quintuplet 
cluster, for which we obtain an upper limit of $\sim 10$ km/s. An accurate measurement
of the internal velocity dispersion will additionally constrain the virial state 
and the long-term dynamical stability of the young 
star cluster in the Galactic center tidal field.

\acknowledgements
The authors sincerely thank the referee for the careful reading of the manuscript,
and for the very positive and helpful referee's report. 
AS, BH and MH acknowledge support by the DFG Emmy Noether programme under grant STO 496/3-1,
and generous travel support by UCLA under NSF grant AST-0909218. AMG and MRM and the work
carried out at UCLA was supported by the NSF under grant AST-0909218 as well. We are grateful 
for numerous discussions with colleagues at both UCLA and the Argelander Institute,
and for the lifely atmosphere at both places and their ongoing scientific support. 
\\

This work is based on observations obtained at the VLT Paranal and Keck, Mauna Kea
observatories, and the authors are most grateful to the support teams of both 
facilities. This work would not have been possible without the intense effort 
and dedication of the Keck LGS-AO staff. We are deeply greatful
for their support enabling these observations. The W. M. Keck Observatory
is operated as a scientific partnership among the California Institute
of Technology, the University of California, and the National Aeronautics
and Space Administraction. The Observatory was made possible by the generous
financial support of the W. M. Keck Foundation. The authors wish to recognize
and acknowledge the very significant cultural role and reverence that the 
summit of Mauna Kea has always had within the indigenous Hawaiian community.
We are most fortunate to have the opportunity to conduct observations from 
this mountain. We also sincerely thank the VLT/NACO staff for their substantial
efforts to provide high-quality observations across our multi-epoch campaigns.

% Table 1 -- observation log

\begin{table*}
\footnotesize
\caption{\label{obstab} Log of Keck/NIRC2 and VLT/NAOS-CONICA Quintuplet imaging observations}
\tiny
\begin{tabular}{lllllcrrrrcc}
\hline
UT Date & Instrument & pixscale & FOV & Filter & $t_{exp}$ [s] & coadds & $N_{obs}$ & $N_{used}$ & $t_{int}$ [s] & FWHM [mas] & Strehl  \\
\hline
2003 July 23  & VLT/NACO   & 27.1 & 27$''$ & $K_s$  & 2.0  & 30 &  16 & 16 & 960 & 75-83 & 7-10\% \\
2003 July 22  & VLT/NACO   & 27.1 & 27$''$ & $K_s$  & 20.0 &  2 &  16 & 16 & 640 & 71-82 & 3-5\% \\
2008 July 24  & VLT/NACO   & 27.1 & 27$''$ & $K_s$  & 2.0  & 15 &  44 & 33 & 990 & 73-81 & 4-12\% \\
2008 May  14  & Keck/NIRC2 & 9.942 & 10$''$ & $K'$     & 1.0  & 30 &  58 & 33 & 990 & 57-70 & 20-34\% \\
2009 May 3-4  & Keck/NIRC2 & 9.942 & 10$''$ & $K'$     & 1.0  & 30 & 117 & 60 & 1800. & 50-62 & 28-47\% \\
\hline
\end{tabular}
\newline
\newline\small
The pixel scale is given in milliarcseconds/pixel, FOV is the field of view, $t_{exp}$ is the exposure time of 
the individual integration, coadds the number of co-added individual exposures in each frame, 
$N_{obs}$ the total number of frames observed, $N_{used}$ the total number of high-quality frames 
entering the deep drizzled image,
and $t_{int}$ is the total integration time in seconds of this deep image. FWHM provides the resolution 
of the deep image after drizzling in milliarcseconds, and Strehl an estimate of the Strehl ratio.
\end{table*}

% Table 2 -- proper motion number counts

\begin{table*}
\caption{\label{pmtab} Proper motion source counts in NACO and NIRC2 data sets}
\begin{tabular}{lll}
\hline
PM data set & Number of sources & notation \\
\hline
NACO 2003 \& NIRC2 2008 \& NIRC2 2009 & 119 & 3-epoch sample \\
NACO 2003 \& NIRC2 2009 & 107 & 2-epoch sample \\
NACO 2003 \& NIRC2 2008 \& 2009 all & 226 & cluster center sample \\
NACO 2003 \& NACO 2008 & 2137 & extended field sample \\
\hline
\end{tabular}
\end{table*}

% Table 3 - NACO-NIRC2 astrometry & photometry (231 matched sources)

\begin{table*}
\caption{\label{nirc2tab} Astrometry \& photometry of NACO 2003 \& NIRC2 2008 \& 2009 sources}
\tiny
\begin{tabular}{cccccccccccccc}
\hline
  Seq & $\delta$RA & $\delta$DEC & $\mu_\alpha\cos\delta$  & $e\mu_\alpha\cos\delta$ & $\mu_\delta$ & $e\mu_\delta$ & Kp2009 & eK2009 & Kp2008 & eK2008 & Ks2003 & eK2003 & $p_{clus}$  \\
      &  ['']  &   [''] &  [mas/yr]& [mas/yr]& [mas/yr]& [mas/yr]&   [mag] &   [mag] &   [mag] &   [mag] &   [mag] &   [mag] &          \\
\hline
  1 & -0.000 &  0.000 & -0.1561 & 0.1853 & -0.4454 & 0.1522 & 9.404 & 0.144 & 9.302 & 0.154 & 9.858 & 0.157 & 0.845 \\
  2 & -0.092 & -2.244 &  0.0557 & 0.3342 &  0.4961 & 0.2287 & 8.412 & 0.143 & 8.252 & 0.154 & 8.914 & 0.163 & 0.840 \\
  3 & -4.317 &  0.541 & -1.0946 & 0.2427 &  0.0388 & 0.2179 & 7.606 & 0.164 & 7.738 & 0.154 & 7.930 & 0.159 & 0.142 \\
  4 &  0.697 &  3.326 &  0.1949 & 0.1281 &  0.4563 & 0.2261 & 9.256 & 0.140 & 9.295 & 0.154 & 9.591 & 0.133 & 0.908 \\
\hline
\end{tabular}
\newline
\newline\small
Positions are given as offsets relative to the Quintuplet proper member Q12 (RA 17:46:15.13, DEC -28:49:35.07). 
The proper motion $\mu_\alpha\cos\delta$ corresponds to the motion in the East-West direction 
($\mu_\alpha\, cos\delta$, with $\alpha$ Right Ascension and $\delta$ Declination), 
$\mu_\delta$ corresponds to the North-South motion of each star. Photometry of all three epochs
is also provided, with photometry of stars with $Ks < 14$ mag taken from the NACO 2003 2s exposures, 
while fainter photometry is supplemented from the deep 20s integrations. 
Columns 5 and 7 contain the proper motion uncertainties in each direction,
and columns 9, 11, and 13 contain the photometric uncertainties in each epoch.
Column 14 provides a membership probability indicator.
Monte Carlo simulations of the proper motion plane for one of the outer cluster fields
(the Pistol field, Field 2 in Hu{\ss}mann 2014) suggest that cluster and field stars are 
most efficiently separated with a formal probability threshold of $p_{cluster} > 0.4$ 
(see Sections 4.2.2.1 to 4.2.2.3 in Hu{\ss}mann (2014) for details).
Table \ref{nirc2tab} is published in its entirety in the electronic edition of ApJ, with the first rows
shown here as guidance regarding its form and content.
\end{table*}

% Table 4 - NACO Astrometry & Photometry full field

\begin{table*}
\caption{\label{nacotab} Astrometry \& photometry of NACO 2003 \& NACO 2008 sources}
\tiny
\begin{tabular}{cccccccccccccc}
\hline
  Seq & $\delta$RA & $\delta$DEC & $\mu_\alpha\cos\delta$  & $e\mu_\alpha\cos\delta$ & $\mu_\delta$ & $e\mu_\delta$ & Ks2003 & eK2003 & Ks2008 & eK2008 & $p_{clus}$  \\
      &  ['']  &   [''] &  [mas/yr]& [mas/yr]& [mas/yr]& [mas/yr]&   [mag] &   [mag] &   [mag] &   [mag] &          \\
\hline
   1  & 10.754 & -3.031 & -0.1680 & 1.0420 &  0.1060 & 0.9900 &  9.359 &  0.139 &  9.040 &  0.015 &  0.303 \\
   2  &  3.622 &  0.268 & -0.2560 & 0.4310 & -0.3700 & 0.5930 &  9.379 &  0.051 &  8.970 &  0.007 &  0.579 \\
   3  & -0.300 &  5.447 & -3.0900 & 0.2830 & -0.6450 & 0.7540 &  9.385 &  0.064 &  9.511 &  0.007 &  0.000 \\
   4  &  0.695 &  3.338 & -0.3330 & 0.2210 &  0.1230 & 0.3020 &  9.415 &  0.024 &  9.278 &  0.003 &  0.838 \\
\hline
\end{tabular}
\newline
\newline\small
As in Table \ref{nirc2tab}, positions are given relative to the Quintuplet proper member Q12 (RA 17:46:15.13, DEC -28:49:35.07). 
The proper motion $\mu_\alpha\cos\delta$ corresponds to the motion in the East-West direction 
($\mu_\alpha\, \cos\delta$, with $\alpha$ Right Ascension and $\delta$ Declination), 
$\mu_\delta$ corresponds to the North-South motion of each star. Photometry is provided for both NACO epochs, 
with photometry of stars with $Ks < 14$ mag taken from the NACO 2003 2s exposures, 
while fainter photometry
is supplemented from the deep 20s integrations. 
Columns 5 and 7 contain the proper motion uncertainties in each direction,
and columns 9 and 11 contain the photometric uncertainties in each epoch.
Column 12 provides a membership probability indicator (see notes to table \ref{nirc2tab} for explanation).
Table \ref{nacotab} is published in its entirety in the electronic edition of ApJ, with the first rows
shown here as guidance regarding its form and content.
\end{table*}

% Table 5 -- movefits

\begin{table*}
\caption{\label{fittab} Fitted parameters of the cluster and field distributions}
\begin{tabular}{lccc}
\hline
Data set      & NIRC2-NACO all &  3-epoch  & NACO-NACO  \\ 
\hline
$N_{stars}$     &   215     &     119  &   1968    \\       
$\mu$ East-West cluster [mas/yr]   & -0.04     &  -0.05   &   0.01   \\   
$\mu$ North-South cluster [mas/yr] &  0.03     &   0.07   &   0.08   \\     
$\mu$ East-West field  [mas/yr]    &  1.61     &   2.09   &   1.69   \\     
$\mu$ North-South field [mas/yr]   & -2.66     &  -3.16   &  -2.32   \\     
separation [mas/yr]                &  3.16     &   3.87   &   2.93   \\  
GP angle   [degrees]               &  33.0     &   32.3   &   32.8   \\  
semi-major axis cluster [mas/yr]  &  0.63     &   0.54   &   0.88   \\  
semi-minor axis cluster [mas/yr]  &  0.61     &   0.49   &   0.62   \\  
semi-major axis field   [mas/yr]  &  5.28     &   4.66   &   5.75   \\  
semi-minor axis field   [mas/yr]  &  2.67     &   2.39   &   2.41   \\  
fraction of cluster stars         &  0.37     &   0.55   &   0.27   \\    
\hline
\end{tabular}
\newline
\newline\small
Columns represent the ($x,y$) position of the fitted cluster and field ellipses
($\mu$ East-West cluster, $\mu$ North-South cluster, $\mu$ East-West field, 
$\mu$ North-South field), the separation of the centroids, 
the angle of the field ellipse indicating the orientation of the Galactic plane, 
the semi-major and -minor axes of the cluster and field ellipses (2-sigma gaussian 
parameters of each fit),
and the fraction of cluster stars relative to the fitting sample size. 
The separation between the fitted cluster and field ellipses denotes the relative motion 
of the Quintuplet cluster with respect to the field.
\end{table*}
\clearpage

%%% Fig. 1 -- nice images

%%% Fig. 2 -- pos uncertainties 

%%% Fig. 3 -- final pm uncertainties

%%% Fig. 4 -- proper motion diagramme with uncertainties
\clearpage

%%% Fig. 5 -- the real motion fits :)

%%% Fig. 6 -- here comes the final fit! :)

%%% Fig. 7 -- the nice colour image first! :)

%%% Fig. 8 -- there be orbits!!!

%%% Fig. 9 -- initial vs. next orbit

%%% Fig. 10 -- min/max elevation above/below the Galactic plane

%%% Fig. 11

%%% Fig. 12

\clearpage
{\sc References}

\small\sl
\noindent
Athanassoula, E. 1992, MNRAS, 259, 345 \\
Bartko, H., Martins, F., Trippe, S.,  Fritz, T. K., et al. 2010, ApJ, 708, 834 \\
Binney, J., Gerhard, O. E., Stark, A. A., et al.~1991, MNRAS, 252, 210 \\
Bishop, C. M. 2006, Pattern Recognition and Machine Learning, (New York, NY: Springer, 1st ed.) \\
B\"oker, T., Falc'on-Barroso, J., Schinnerer, E., Knapen, J. H., Ryder, S. 2008, AJ, 135, 479 \\
Clarkson, W., Ghez, A. M., Morris, M., Lu, J. R., McCrady, N., Stolte, A., Do, T., Yelda, S.~2012, ApJ, 751, 132 \\
Cotera, A. S., Erickson, E. F., Colgan, S. W. J., Simpson, J. P., Allen, D. A., Burton, M. G. 1996, ApJ, 461, 750 \\
Crowther, P. A. 2007, ARA\&A, 45, 177  \\
Dame, T. M., Hartmann, D., Thaddeus, P. 2001, ApJ, 547, 792 \\
Diolaiti, E., Bendinelli, O., Bonaccini, D., et al. 2000, A\&AS, 147, 335 \\
Do, T., Ghez, A. M., Morris, M. R., Lu, J. R., Matthews, K., Yelda, S., Larkin, J. 2009, ApJ, 703, 1323 \\
Do, T., Lu, J. R., Ghez, A. M., Morris, M. R., Yelda, S., Martinez, G. D., Wright, S. A., Matthews, K. 2013, ApJ, 764, 154 \\
Dong, H., Wang, Q. D., Cotera, A., Stolovy, S., Morris, M. R., Mauerhan, J., Mills, E. A., Schneider, G., Calzetti, D., Lang, C. 2011, MNRAS, 417, 114 \\
Elmegreen, D. M., Elmegreen, B. G. 1978, ApJ, 219, 105 \\
Espinoza, P., Selman, F. J., Melnick, J. 2009, A\&A, 501, 563 \\
Figer, D. F., Kim, S. S., Morris, M., Serabyn, E., Rich, R. M.,  McLean, I. S. 1999a, ApJ, 525, 750 \\
Figer, D. F., McLean, I. S., Morris, M. 1999b, ApJ, 514, 202 \\
Fruchter, A. S., Hook, R. N. 2002, PASP, 114, 144 \\
Fujii, M., Iwasawa, M., Funato, Y., Makino, J. 2008, ApJ, 686, 1082 \\
Gardner, E., Flynn, C. 2010, MNRAS, 405, 545 \\
\\
Genzel, R., Sch\"odel, R., Ott, T., Eisenhauer, F., et al. 2003, ApJ, 594, 812 \\
Gerhard, O. 2001, ApJL, 546, 39 \\
Ghez, A. M., Salim, S., Hornstein, S. D., Tanner, A., Lu, J. R., Morris, M., Becklin, E. E., Duchêne, G. 2005, ApJ, 620, 744 \\
Glass, I. S., Moneti, A., Moorwood, A. F. M. 1990, MNRAS, 242, 55 \\
Habibi, M., Stolte, A., Brandner, W., Hu{\ss}mann, B., Motohara, K. 2013, A\&A, 556, A26 \\
Habibi, M., Stolte, A., Harfst, S. 2014, A\&A, 566, A6 \\
Hu{\ss}mann, B., Stolte, A., Brandner, W., Gennaro, M., Liermann, A. 2012, A\&A, 540, A57 \\
Hu{\ss}mann, B., 2014, PhD Thesis, University of Bonn, http://hss.ulb.uni-bonn.de/2014/3485/3485.htm \\
Kim, S. S., Figer, D. F., Lee, H. M., Morris, M. 2000, ApJ, 545, 301  \\
Kim, S. S.,  Morris, M. 2003, ApJ, 597, 312 \\
Kim, S. S., Saitoh, T. R., Jeon, M., Figer, D. F., Merritt, D., Wada, K. 2011, ApJ {\sl Letters}, 735, 11 \\
Kozhurina-Platais, V., Girard, T. M., Platais, I., van Altena, W. F., Ianna, P. A., Cannon, R. D. 1995, AJ, 109, 672 \\
Lang, C. C., Cyganowski, C., Goss, W. M., Zhao, J.-H. 2003, ANS, 324, 1 \\
Launhardt, R., Zylka, R., Mezger, P. G. 2002, A\&A, 384, 112 \\
Lawrence, A., Warren, S. J., Almaini, O., Edge, A. C., et al. 2007, MNRAS, 379, 1599 \\
Lenzen, R., Hartung, M., Brandner, W., Finger, G., et al. 2003, Proc. SPIE, 4841, 944 \\
Liermann, A., Hamann, W.-R., Oskinova, L. M. 2009, A\&A, 494, 1137 \\
Liermann, A., Hamann, W.-R., Oskinova, L. M., Todt, H., Butler, K. 2010, A\&A, 524, A82 \\
Liermann, A., Hamann, W.-R., Oskinova, L. M. 2012, A\&A, 540, A14 \\
Lindqvist, M, Winnberg, A., Habing, H.J., Matthews, H.E. 1992, A\&AS, 92, 43 \\
Longmore, S. N., Rathborne, J., Bastian, N., et al. 2012, ApJ, 746, 117 \\
Longmore, S. N., Kruijssen, J. M. D., Bally, J., et al. 2013, MNRAS Letters, 433, 15 \\
Lu, J. R., Ghez, A. M., Hornstein, S. D., Morris, M. R., Becklin, E. E., Matthews, K. 2009, ApJ, 690, 1463 \\
Lu, J. R., Do, T., Ghez, A. M., Morris, M. R., Yelda, S., Matthews, K. 2013, ApJ, 764, 155 \\
Lucas, P. W., Hoare, M. G., Longmore, A., Schr\"oder, A. C., et al. 2008, MNRAS, 391, 136 \\
Mackey, J., Lim, A. J. 2010, MNRAS, 403, 714 \\
Martins, F., Hillier, D. J., Paumard, T., Eisenhauer, F., Ott, T., Genzel, R. 2008, A\&A, 478, 219 \\
Mazzuca, L. M., Knapen, J. H., Veilleux, S., Regan, M. W. 2008, ApJS, 174, 337 \\
Molinari, S., Bally, J., Noriega-Crespo, A., et al. 2011, ApJ Letters, 735, 33 \\
Mauerhan, J. C., Morris, M. R., Cotera, A., Dong, H., et al. 2010a, ApJL, 713, 33  \\
Mauerhan, J. C., Cotera, A., Dong, H., Morris, M. R., et al. 2010b, ApJ, 725, 188 \\
Nagata, T., Woodward, C. E., Shure, M., et al. 1990, ApJ, 351, 83 \\
Nagata, T., Woodward, C. E., Shure, M., Kobayashi, N. 1995, AJ, 109, 1676 \\
Namekata, D., Habe, A., Matsui, H., Saitoh, T. 2009, ApJ, 691, 1525 \\
Oka, T., Hasegawa, T., Hayashi, M., et al. 1998, ApJ, 493, 730 \\
Okuda, H., Shibai, H., Nakagawa, T., et al. 1990, ApJ, 351, 89 \\
Paumard, T., Genzel, R., Martins, F., Nayakshin, S., et al. 2006, ApJ, 643, 1011 \\
Portegies Zwart, S. F., Makino, J., McMillan, S. L. W., Hut, P. 2002, ApJ, 565, 265 \\
Press, W. H., Teukolsky, S. A., Vetterling, W. T., Flannery, B. P. 2007, Numerical recipes in C++ : the art of scientific computing, (Cambridge University Press; 3rd ed.), Chpt. 16, pp. 842 \\
Regan, M. W., Teuben, P. 2003, ApJ, 582, 723 \\
Rochau, B., Brandner, W., Stolte, A., Gennaro, M., et al. 2010, ApJL, 716, 90 \\
Rodriguez-Fernandez, N. J., Combes, F. 2008, A\&A, 489, 115 \\
Rousset, G., Lacombe, F., Puget, P., Hubert, N. N., et al. 2003, Proc. SPIE, 4839, 140 \\
Sch\"odel, R., Eckart, A., Alexander, T., Merritt, D., et al.~2007, A\&A, 469, 125 \\
Sch\"odel, R., Najarro, F., Muzic, K., Eckart, A. 2010, A\&A, 511, A18 \\
Stolte, A., Ghez, A. M., Morris, M., Lu, J. R., Brandner, W., Matthews, K. 2008, ApJ, 675, 1278 \\
Sumi, T., Eyer, L., Wo\'zniak, P. R. 2003, MNRAS, 340, 1346 \\
Tuthill, P., Monnier, J., Tanner, A., Figer, D., Ghez, A., Danchi, W. 2006, Science, 313, 935 \\
Wang, Q. D., Dong, H., Cotera, A., Stolovy, S., Morris, M., Lang, C. C., Muno, M. P., Schneider, G., Calzetti, D. 2010, MNRAS, 402, 895 \\
Warren, S. J., Hambly, N. C., Dye, S., Almaini, O., et al. 2007, MNRAS, 375, 213 \\
Yelda, S., Lu, J. R., Ghez, A. M., Clarkson, W., Anderson, J., Do, T., Matthews, K. 2010, ApJ, 725, 331 \\

\end{document}